\documentclass[usenatbib]{mnras}
\usepackage[T1]{fontenc}
\DeclareRobustCommand{\VAN}[3]{#2}
\let\VANthebibliography\thebibliography
\def\thebibliography{\DeclareRobustCommand{\VAN}[3]{##3}\VANthebibliography}
\usepackage{amsmath}	% Advanced maths commands
\usepackage{amssymb}	% Extra maths symbols
\usepackage{comment}
\usepackage{array}
\usepackage{tabularx}
\usepackage{float}
\usepackage{bigints}
\usepackage{mathtools}
\usepackage{xfrac}
\usepackage[dvipsnames]{xcolor}
\usepackage{multicol} %insert image in full width
\usepackage{multirow}
\usepackage{subcaption} %combine subfigures
\usepackage{arydshln}
\usepackage{booktabs}
\usepackage{tabularx,colortbl}

\newcommand{\bc}{\begin{comment}}
\newcommand{\ec}{\end{comment}}

\newcommand{\mpch}{${h^{-1}\rm Mpc}$~}
\newcommand{\mpcH}{${h^{-1}\rm Mpc}$}

\newcommand{\esD}{$\Delta\Sigma \mkern6mu $}
\newcommand{\vcutsymbol}{{\ooalign{\hfil$\vee$\hfil\cr\kern0.08em--\hfil\cr}}}

\newcommand\Tstrut{\rule{0pt}{4.0ex}}         % = `top' strut
\newcommand\Bstrut{\rule[-3ex]{0pt}{0pt}}   % = `bottom' strut

\title[Subaru HSC weak lensing  of redMaPPer satellites]{Subaru HSC weak lensing of SDSS redMaPPer cluster satellite galaxies: Empirical upper limit on orphan fractions}
\author[ Amit Kumar et al.]{
Amit. Kumar,$^{1}$\thanks{E-mail: amitk@iucaa.in (IUCAA)}
Surhud More,$^{1, 2}$\thanks{E-mail: surhud@iucaa.in (IUCAA)}
Divya Rana$^{1}$
\\
$^{1}$ Inter-University Centre for Astronomy and Astrophysics, Post bag 4, Ganeshkhind, Pune 411007, India\\
$^{2}$ Kavli Institute for the Physics and Mathematics of the Universe (WPI), 5-1-5 Kashiwanoha, 2778583, Japan
}
\date{Accepted XXX. Received YYY; in original form ZZZ}
\pubyear{2021}
\begin{document}
\label{firstpage}
\pagerange{\pageref{firstpage}--\pageref{lastpage}}
\maketitle

\begin{abstract}
Weak gravitational lensing directly probes the matter distribution surrounding satellite galaxies in galaxy clusters. We measure the weak lensing signal induced on the shapes of background galaxies around SDSS redMaPPer cluster satellite galaxies, which have their central galaxies assigned with a probability $P_{\rm cen}>0.95$ in the redshift range, $0.1\leq z\leq 0.33$. We use the galaxy shapes from the Subaru Hyper Suprime-Cam (HSC) survey for this purpose. We bin satellite galaxies by their distance from the cluster centre and compare it to the signal around a control sample of galaxies which do not reside in clusters but have similar colours and magnitudes. We explore the effect of environmental processes on the dark matter mass around satellites. We see hints of a difference in the mass of the subhalo of the satellite compared to the halo masses of galaxies in our control sample, especially in the innermost cluster-centric radial bin ($0.1<r<0.3$ [\mpcH]). For the first time, we put an upper limit on the prevalence of orphan galaxies which have entirely lost their dark matter halos with cluster-centric distances with the help of our measurements. However, these upper limits could be relaxed if there is substantial contamination in the satellite galaxy sample.
\end{abstract}


\section{Introduction}
Satellite galaxies within galaxy clusters live in a dense environment, where their baryonic and dark matter components are susceptible to environmental effects. In contrast, field galaxies behave more like 'island Universes' \citep{Tormen_1998, Tormen2004, Lucia2004, Gao2004}. Various processes can affect the characteristics of galaxies in dense galaxy clusters. Ram-pressure stripping  can remove the hot gas, hence quench the star formation of a galaxy \citep{Gunn1972}. A satellite galaxy may disrupt its spiral arms and turn into an elliptical galaxy due to frequent encounters with other satellite galaxies \citep{Moore1996, Moore1998} in a process called galaxy harassment. In addition galaxies can undergo mergers in these dense environments, where survival may result in single massive galaxy at the cluster center \citep{Merritt1985, Bosch2005, Corney2007}. Furthermore, gravitational interactions can tidally strip the dark matter subhalos and dictate the evolution of  galaxies falling into the cluster \citep{ Merritt83,Giocoli2008}. 

The tidal effects are stronger near the cluster center (i.e., near the brightest cluster galaxy, hereafter BCG), and they fall off as a function of the distance of the satellite from the BCG \citep{Springel2001, Lucia2004, Gao2004, Zhao2004, Xie2015}. Simulations suggest preferential stripping of dark matter up-to a large extent prior to the baryonic matter due to tidal forces \citep{Smith2016}. The satellite galaxies also feel dynamical friction because of their orbital motion around the BCG. Collisionless dark matter particles transfer energy and momentum to the particles of the host during such motion. Consequently, the orbit of the satellite decays with time, transporting it towards the center of the host's potential well \citep{Binney2008}. In addition, self-interactions between dark matter particles \citep{Spergel2000} can cause the dark matter in subhalos to evaporate \citep[][]{Bhattacharyya_2021}. Thus, there are interesting physical processes that can affect the dark matter distribution around satellite galaxies, and we expect satellites and their subhalos to evolve differently than galaxies residing in non-cluster environment. Observations of dark matter distribution around satellite galaxies and their comparison with dark matter distribution around similar field galaxies can be used to constrain the efficiency of such environmental processes. 

Cosmological simulations can be used to study the environmental processes that affect satellite galaxies in galaxy clusters. Collisionless simulations have highlighted mass loss due to tidal stripping for subhalos in galaxy clusters purely due to gravitational effects \citep{Ghinga1998, Tormen2004, Gao2004, Bosch2005, Nagai2005, Giocoli2008, Xie2015, Rhee2017}. However there is growing evidence that numerical simulations can artificially disrupt subhalos due to insufficient mass and force resolution \citep[see e.g.,][]{vandenBosch2017, vandenBosch2018, vandenBosch2018b}. Studies that simultaneously aim to fit the abundance and clustering of galaxies and use subhalos in simulations as hosts of satellite galaxies thus need to invoke substantial fractions of {\it orphan galaxies}, galaxies that have lost their subhalos entirely in the simulations \citep[see e.g.,][]{Behroozi_2019}.

Furthermore, the stripping and disruption of subhalos can be affected by the presence of baryons. Hydrodynamical simulations can also capture the effects of the environment on the baryons within satellite galaxies, however they also suffer from limited numerical resolution \citep{Vogelsberger2014}. The goal of this paper is to find observational evidence for the combined environmental effects on the dark matter subhalo masses of these satellite galaxies.

The evolution of satellite and field galaxies can be studied using their stellar-to-halo mass relation (hereafter SHMR). Weak gravitational lensing provides a way to probe the mass distribution around these galaxies. A number of studies have inferred the SHMR for central galaxies with a variety of methods including, satellite kinematics, strong and weak lensing observations \citep{Hoekstra2005, heymans2006, Mandelbaum2006, More2009, Mandelbaum2016, More2011, van_uitert2011, van_uitert2016, Leauthaud2012, valander2014, zu2015, Coupon2015}. Measurement of the SHMR for satellite galaxies is challenging. \cite{Natarajan2009}, used images from Hubble Space Telescope to measure the subhalo masses at different radial separations, for $L_*$ galaxies from C10024+16(z=0.39) galaxy cluster, with the combination of strong and weak gravitational lensing. \cite{Okabe2014} measured the weak lensing signal around the nearby Coma cluster, and computed masses of 32 subhalos from the cluster by constructing shear map using 4 deg$^2$ deep field imaging data from Suprime-cam from the Subaru telescope. To obtain results for a larger statistical sample, one has to resort to stacking the weak lensing signal around satellite galaxies \citep[see e.g.,][]{Yang2006, Poster_mira2011}.

We are in the era of large imaging surveys, which have made such studies possible. \citet{Li2014} used the stacked galaxy-galaxy lensing measurements to constrain the subhalo masses ($\rm \log(M_{sub}/h^{-1} M_{\odot}) = 11.68 \pm 0.76$) in massive galaxy groups $10^{13}$-\;$5\times10^{14} h^{-1} \rm   M_{\odot} $ with the CHFT lensing survey \citep{Heymans2012}. On galaxy group scales, \citet{Sifon2015} presented constraints on the subhalo masses of satellite galaxies in a larger sample of more than 10000 spectroscopically selected GAMA galaxy groups \citep{Robotham2011}, using the 100 deg$^2$ of Kilo-Degree lensing survey \citep{de_Jong2013}. On galaxy cluster scales, \citet{Li2016} used the CFHT Stripe 82 survey \citep{CH82} to measure the subhalo to stellar mass ratio for redMaPPer satellites in different projected radial distance bins. In addition, \citet{Sifon2018} have also carried out a detailed analysis of the stellar mass-subhalo mass relation of satellite galaxies in MENeaCS, and its dependence upon cluster-centric distance with a massive cluster sample \citep{Sand2012}, using imaging data from the Canada-France-Hawaii Telescope.

In this paper, we carry out a similar analysis as \citet{Li2016}, but utilizing data from the first year galaxy shape catalog from the Subaru HSC-SSP(see:\ref{sec:hsc}) survey. Ideally one would like to track the subhalo masses of galaxies as they infall in to a cluster halo to their current position. However such information is inaccessible to us from observations. Therefore, we use a control sample of galaxies in the field as a proxy of galaxies that have not entered the cluster halos. It is expected that the photometric properties of the satellite galaxies can change between the infall on to the cluster and the current position of the galaxy and that a control sample should ideally account for such differences. However, \citet{WatsonConroy2013} studied the differences in the stellar mass and (sub)-halo mass between central and satellite galaxies\footnote{The subhalo mass is defined at accretion in this case.}, and found approximately a 10 percent difference between the two. A large part of this difference can also be explained by the fact that halos undergo mass growth due to pseudo-evolution \citep[see fig. 10 in][]{More_2015}, given that the subhalo mass at accretion is defined at an earlier point in time \citep{Diemer2013}. Given that these differences are not vast, we will compare the subhalo masses of satellite galaxies selected from an optical cluster catalog to the halo masses of a control sample of galaxies selected to have similar photometric properties as the satellites. This ensures that the control sample of galaxies have the same current baryonic properties as well as redshift distributions as our satellite galaxy sample.

This paper is structured as follows. We describe the data we use in Section~\ref{sec:data}, followed by the procedure we follow to measure the weak lensing signal in Section~\ref{sec:weak_lensing_theory}. We define our model for the signal around the satellite galaxies and our control sample of galaxies in Section~\ref{sec:modelling}. We present the statistical error estimates on the weak lensing signal and the methodology we adopt for fitting our model to the data in Section~\ref{sec:stastical_estimates_and_model_fits}. In Section~\ref{sec:results_and_discussion}, we discuss our results and their implications followed by a summary of our results in Section~\ref{sec:summary}.

In this paper, we use the terms "halo" and "subhalo" to refer to the dark matter halos corresponding to central and satellite galaxies, respectively. We will refer to the three dimensional radial distance from the center of the satellite galaxy as "r" and its projection on the sky as "R". Throughout this analysis, we use flat $\Lambda$CDM cosmology with parameters: $\Omega_m= 0.315, \Omega_bh^2=0.02205, \sigma_8=0.829, n_s=0.9603$ \citep{Plank2014}, and distances in comoving \mpch units.
\section{Data}
\label{sec:data}
In this section, we describe the data we use to carry out the weak lensing analysis of satellite galaxies in galaxy clusters. In particular, as described in Section~\ref{sec:redmapper}, we use the galaxy cluster members identified in the SDSS redMaPPer cluster catalog as lens galaxies, while we use galaxies identified from the first year data of the HSC-SSP survey as background sources, as described in Section~\ref{sec:hsc}. 

\subsection{The redMaPPer cluster catalogue} 
\label{sec:redmapper}
The red-sequence photometric cluster finding algorithm (redMaPPer) identifies galaxy clusters as overdensities of red-sequence galaxies in large imaging surveys \citep{Rykoff2014,Rykoff2016}. The algorithm is designed for optimal use of current and upcoming large photometric surveys. Briefly, it constructs a model for the red sequence of galaxies as a function of redshift with the help of a nominal training sample of spectroscopic galaxies. The algorithm self-calibrates this model iteratively using multi-band photometric data, which ultimately results in photometric redshift estimates ($\sigma_z/(1+z) \approx 0.01$ for z $\leq$ 0.7) for galaxy clusters \citep{Rykoff2016}. Cluster membership probabilities are assigned to satellite galaxies, which are then further used to define the richness of the cluster ($\lambda$). When compared to samples with spectroscopic redshifts, the memberships are accurate to 1\% and show low (<5\%) projection effects according to \citet{Rozo_2014}. Each cluster is assigned a photometric redshift ($z_{\rm \lambda}$) based on the high probability cluster members. One of the salient features of the redMaPPer algorithm is that it assigns a probability, $P_{\rm cen}$ to five of the members in a galaxy cluster to correspond to its center based on the overall radial, luminosity, and redshift distribution of member galaxies. 

\begin{figure*}
    \centering
    \includegraphics[width=\textwidth]{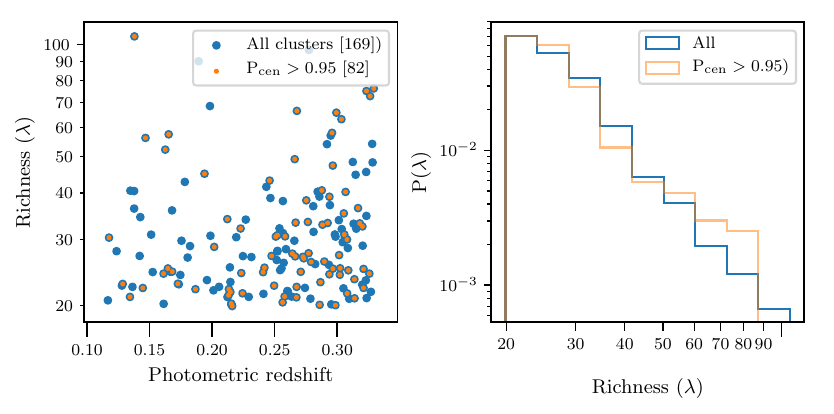}
    \caption{The left hand panel shows a scatter plot of all redMaPPer clusters in the HSC survey footprint in blue symbols. The red symbols correspond to the clusters selected after applying the cut of $P_{\rm cen}$>0.95. The right hand panel shows that the richness distribution of the entire cluster sample is largely unchanged after we apply the cut on $P_{\rm cen}$.}
    \label{fig:pz_lambda}
\end{figure*}

In our analysis, we use galaxy clusters and randoms from the redMaPPer catalog v6.3 \citep{Rykoff2016} obtained from the SDSS DR8 \citep{Aihara2011} photometric galaxy catalog. We only use those clusters and randoms which overlap  with the HSC year-1 shape catalog footprint. We also restrict our lens sample to a redshift range $0.1\leq z\leq 0.33$ where the incompleteness corrections due to the flux limit of SDSS are small. We have 169 such clusters in the HSC footprint. Furthermore, to avoid issues related to miscentering, we only choose satellites from galaxy clusters where the most probable central galaxy has a centering probability $P_{\rm cen}$>0.95. This cut finally leaves us with a sample of 82 clusters as can be seen from the left hand panel of Fig.~\ref{fig:pz_lambda}. The right hand panel shows that the cut on $P_{\rm cen}$ has a relatively minor effect on the richness distribution of clusters in our sample.

\subsection{ The HSC-SSP Survey}
\label{sec:hsc}
The Hyper Suprime-Cam Subaru Strategic Program \citep{aihara2017,Aihara2018,aihara2019} is an ongoing multi-band (grizy plus 4 narrow-band filters) imaging survey, that is being conducted with the Hyper Suprime-Cam instrument on the 8.2m Subaru Telescope \citep{miyazaki2012, Miyazaki_2015, miyazaki2017, komiyama2017, furusawa2017, kawamoto2018}. The HSC instrument is a 0.9 Gigapixel camera with a $0.168$ arcsec pixel scale, and it consists of 104 science CCDs which cover a 1.5 degree diameter field of view. The combination of a wide field of view, a large aperture, and an excellent site with good seeing conditions (median of 0.6\arcsec in the $i$-band), makes HSC a unique instrument to carry out weak lensing studies. In our study we use the shape catalog of galaxies from the Subaru Hyper Suprime-Cam survey.

The HSC first year shape catalog covers nearly $\sim $136 deg$^2$  sky area in six disjoint fields (XMM, GAMA09H, WIDE12H, GAMA15H, VVDS, \& HECTOMAP), observed over 90 nights in total, during March 2014 to Nov 2015, described exhaustively in  \citet{Mandelbaum2018a} and \citet{Mandelbaum2018b}. The survey fields are chosen such that the HSC footprint entirely overlaps with SDSS's Baryon Oscillation Spectroscopic Survey \citep[BOSS;][]{Dawson2013} which provides a large well characterized spectroscopic sample of galaxies with z$\sim$0.7 which can act as lens galaxies and also be used to train and calibrate photometric redshift finding algorithms. It also allows the HSC survey to perform joint cosmological analysis with galaxy-galaxy lensing and galaxy clustering \citep{Miyatake_2021,Sugiyama_2021}, in addition to training photometric redshift algorithms \citep{Tanaka2018}. In our work, we use the shape catalogue from the data release S16a \citep{aihara2019}, which is an incremental release of the public data release from the HSC Survey \citep{Aihara2018}.

\cite{Mandelbaum2018a} implemented a number of conservative cuts to select galaxies in the shape catalog in order to meet the requirements for carrying out weak lensing analysis. For example, the full depth and full color cut (FDFC cut) restricts galaxies to those sky regions which reach the nominal full depth in all five band filters (g,r,i,z,y). This ensures the uniformity of the sample and better determination of photometric redshifts. \cite{Mandelbaum2018a} also restrict galaxies to have an $i-$band c-model magnitude brighter than i$_{\rm cmodel}$ <  24.5. This is much more conservative than the i-band limiting magnitude of 26.4 \citep[which corresponds to a S/N $\sim~5\sigma$ for PSF photometry, see][]{Aihara2018}.

The shapes of galaxies are measured using a moment-based shape measurement algorithm which utilizes the re-Gaussianization PSF correction method \citep{Hirata2003}. The shapes are measured on coadded images in the i-band with an average seeing FWHM of 0.58\arcsec \citep{Mandelbaum2018a}. The ellipticities are measured as,
    \begin{equation}
    (e_{\rm 1},e_{\rm 2}) = \frac{1-(b/a)^2}{1+(b/a)^2}(\cos{2\psi},\sin{2\psi}) \,.
    \label{eq:e1e2}
    \end{equation}
Here $b/a$ and $\psi$ represent the axis ratio and the angle made by the major axis of the source galaxy with respect to the equatorial coordinate system, respectively. 
\cite{Mandelbaum2018b} performed detailed image simulations of galaxies using the {\sc GalSim}  \footnote{https://github.com/GalSim-developers/GalSim} \citep{Rowe2015} accounting for the survey properties such as the depth and the seeing of the HSC survey in order to calibrate the multiplicative (m) and additive biases ($c_{\rm 1},c_{\rm 2}$), 
and other quantities such as the RMS intrinsic distortions  e$_{\rm rms}$ and the photon noise per component in galaxy shapes $\sigma_{\rm e}$, required to assign optimal weights (Eq.~\ref{eq:shape_weight}) to the galaxy shape measurements. 
The shape weight is related to the RMS intrinsic dispersion $e_{\rm RMS}$ and $\sigma_{\rm e}$ by,
\begin{equation}
w_{\rm s}=\frac{1}{(\sigma_{\rm e}^2+e_{\rm RMS}^2)}\,.
\label{eq:shape_weight}
\end{equation}

For overall shear calibrations, \cite{Mandelbaum2018b}, estimate the residual systematic uncertainties to be of the order 0.01.
Extensive null tests of the shear catalog are presented in \cite{Mandelbaum2018a}, \cite{Oguri2018}, and \cite{hikage2018} to demonstrate that the systematic uncertainties in the shear  catalog are small enough that the cosmological weak lensing  analyses using galaxy-galaxy lensing and cosmic shear will be predominantly dominated by statistical error. Even after relatively considerate cuts, the HSC first-year shear catalog produces galaxy shapes with a high source number density 24.6 (raw), and 21.8 (effective) arcmin$^{-2}$ \citep{Mandelbaum2018a} at a median redshift of $0.8$, which makes it ideal to study the weak lensing of the redMapper satellite galaxies which we restrict to redshifts less than $0.33$. We note that the systematic effects on shear recovery in dense fields has not yet been properly quantified for the HSC survey. Such characterizations would require detailed image simulations around cluster regions, an area of focus for our near future work, but it is beyond the scope of the current paper.

\subsubsection{HSC Photometric redshift catalog }
\label{sec:hsc_photometric_catlog}
Photometric redshift uncertainties are one of the major source of systematics in interpreting the weak lensing signal measured around the lenses. Photometric redshifts are used to identify the source galaxies which lie behind the lens galaxies in order to measure the weak lensing signal. Inclusion of foreground galaxies as a part of background sources due to incorrect determination of photometric redshifts can dilute the lensing signal and lead to an underestimate of the mass profile \citep{Broadhurst2005}. However, acquiring spectroscopic redshifts for all the source galaxies is virtually impossible for current and upcoming  weak lensing surveys due to the faint magnitude limits as well as the high source number density of galaxies. Therefore photometric redshifts are inevitable. 

Multi-wavelength observations of galaxies help to sample the spectral energy distribution of the galaxies in broad bands and thus are essential for the accurate determination of redshifts. All on-going weak lensing surveys, e.g. Kilo-Degree Survey \citep[KIDS,][]{Kuijken2019}, Dark Energy Survey \citep[DES,][]{Abbott2022}, Hyper Suprime-Cam \citep[HSC,][]{Aihara2022}, observe the sky in multiple filters in order to aid in the estimates of photometric redshifts. The depth and the 5-band $(g, r, i ,z, y)$ photometry in the HSC survey allow a reasonably accurate determination of the redshifts of galaxies (with $\sigma_{\rm z}/(1+z) \approx 0.05$ and a 15\% outlier rate for galaxies down to i=25 and 0.04 with 8\% outliers for i<24) over a wide redshift range ($0.2\leq z\leq 1.5$) \citep{Tanaka2018}. 
The HSC survey provides photometric redshifts obtained using 6 different codes. In this work, we use the photometric redshift distributions from the template fitting code {\sc Mizuki} \citep{Tanaka2015}. For each galaxy, {\sc Mizuki} provides the posterior distribution of its redshift by using the multi-band photometry in HSC coupled with physical Bayesian priors. Although the photometric redshift performance is slightly worse for Mizuki compared to other photometric redshift codes (22\% outlier rate for the WL sample and a $\sigma_{\rm z}/(1+z) \approx 0.08$), the template fits allow us a measure of the stellar mass of the galaxies. As we explain in the next section, we also use quite conservative cuts based on the photometric redshift PDF of each galaxy which helps to reduce contamination of the weak lensing signal due to galaxies physically associated with the clusters. We have also verified that our results and conclusions remain similar within our quoted statistical uncertainties even if we use alternative photometric redshifts measured from different algorithms.

\section{Weak gravitational lensing: Theory and measurements}
\label{sec:weak_lensing_theory}

Weak gravitational lensing causes a coherent distortion in the shapes of background source galaxies as a result of the presence of matter between the source and the observer along the line-of-sight. This distortion caused by the lensing shear can be quantified by measuring the observed ellipticities of source galaxies. As galaxies are not intrinsically circular, it is impossible to infer the lensing shear individually from the measured shape of each galaxy. However, it is possible to statistically observe the effect of lensing shear by averaging the shapes of background galaxies that lie a certain projected distance away from the lens galaxy.

Starting from the observed images of the source galaxies (Eq.~\ref{eq:e1e2}), the tangential shape distortion of a source galaxy is given by, 
\begin{equation}
e_{\rm t}= -e_{\rm 1} \cos(2\phi) - e_{\rm 2} \sin(2\phi)\,.
\label{eq:e_t}
\end{equation}
Here $\phi$ is the angle made by the line joining the source to the lens in projected space with respect to the x-axis of the equatorial coordinate system. In the weak lensing regime, the expectation value of the tangential ellipticity at a given distance from the center of the lens is proportional to the tangential shear at that distance \citep{Schramm_Kayser_1995, Seitz_Schneider_1997}. The dominant contribution to the noise comes from the intrinsic non-circular shapes of galaxies, and this noise decreases as $\propto 1/ \sqrt{N}$, where $N$ is the number of lens-source galaxy pairs in a given radial bin.

The gravitational lensing shear is sensitive to the matter density distribution projected along the line-of-sight called the surface mass density, $\Sigma$(R), as well as the anuglar diameter distances between the source, the lens and the observer. For a source galaxy located at redshift z$_{\rm s}$, the tangential component of the shear induced by a lensing object at redshift  z$_{\rm l}$, separated by a comoving projected distance $R$, is given by \citep{Schneider2006},
\begin{equation}
\gamma_{\rm t}(R)=\frac{\Sigma(<R) -\Sigma(R)}{\Sigma_{\rm c}(z_{\rm l},z_{\rm s})}=\frac{\Delta\Sigma(R)}{\Sigma_{\rm c}(z_{\rm l},z_{\rm s})}\,.
\label{eq:esd}
\end{equation}
Here $\Sigma(<R)$ is the average surface mass density within a projected distance $R$, and $\Sigma(R)$ denotes the azimuthally averaged surface mass density within a thin annulus at that distance $R$. The critical surface mass density, $\Sigma_{\rm c}(z_{\rm l},z_{\rm s})$, depends upon the geometry of the lensing configuration, and is related to the angular diameter distances to the lens $D_{\rm A}(z_{\rm l})$, the source $D_{\rm A}(z_{\rm s})$ and between the lens and source $D_{\rm A}(z_{\rm l},z_{\rm s})$ and is given by,
\begin{equation}
\Sigma_{\rm c}(z_{\rm l},z_{\rm s})= \frac{c^2}{4\pi G}\frac{D_{\rm A}(z_{\rm s})}{D_{\rm A}(z_{\rm l}) D_{\rm A}(z_{\rm l},z_{\rm s})(1+z_{\rm l}^2)}\,.
\label{eq:sigma_crit}
\end{equation}
The presence of an extra factor of  $(1+z_{\rm l})^2$ in the denominator is due to our use of comoving coordinates \citep{Bartelmann_and_Schneider_2001}.
Given that the HSC shape catalog provides information about the measured ellipticity and the shape weight i.e.  $e_{\rm 1},e_{\rm 2},w_{\rm s}$ for each source galaxy, along with additive and multiplicative biases, i.e. $c_{\rm 1},c_{\rm 2},m$ in them, we can write the weighted signal, for many-many source-lens galaxy pairs with the help of Eq.~\ref{eq:esd} as,
\begin{equation}
\begin{split}
    \Delta\Sigma(R)&=\frac{1}{(1+\hat{m})}\left(\frac{\sum_{\rm ls}w_{\rm ls} e_{\rm t,ls}\left\langle\Sigma_{\rm c}^{-1}\right\rangle^{-1}}{2\mathcal{R}\sum_{\rm ls}w_{\rm ls} }\right.\\&\qquad\qquad\qquad\qquad\qquad\left.- \frac{\sum_{\rm ls}w_{\rm ls} c_{\rm t,ls}\left\langle\Sigma_{\rm c}^{-1}\right\rangle^{-1}}{\sum_{\rm ls}w_{\rm ls} } \right)\,.
\end{split}    
\label{eq:stacked_signal}
\end{equation}
Here, we use the probability distribution for the source redshifts $p(z_{\rm s})$ to define $\langle\Sigma_{{\rm c}}^{-1}\rangle$ as
\begin{equation}
    \langle\Sigma_{\rm c}^{-1}\rangle=\frac{4\pi G(1+z_{\rm l})^2}{c^2}\int_{z_{\rm l}}^{\infty}\frac{D_{\rm A}(z_{\rm l}) D_{\rm A}(z_{\rm l},z_{\rm s})}{D_{\rm A}(z_{\rm s})} p(z_{\rm s})dz_{\rm s} \,,
\end{equation}
and we conservatively choose only those sources that have an integrated probability to lie at a redshift greater than that of our entire sample of lenses, i.e. 
\begin{equation}
\int_{z_{\rm min}}^{\infty} p(z_{\rm s})dz_{\rm s}>0.99\,,   
\label{eq:photoz}
\end{equation}
where we set $z_{\rm min}$ equals to the maximum redshift of our lens sample ($z_{\rm l_{max}}$ = 0.33) plus some offset factor ($z_{\rm diff}$ = 0.1). Also $w_{\rm ls}=w_{\rm s}\langle\Sigma_{\rm crit}^{-1}\rangle^2$ assigns less weight to the lens-source galaxy pairs closely separated in redshift space, to reduce the contamination of galaxies associated with the lens galaxies in our source galaxy sample. Whereas, $\mathcal{R}$ in the Eq.~\ref{eq:stacked_signal} denotes the shear responsivity, which quantifies how application of small shear affects measurement of $e_{\rm 1},e_{\rm 2}$. For the  observed ellipticity definition used  in Eq.~\ref{eq:e1e2}, the quantity 
\begin{equation}
    \mathcal{R}=1- \frac{\sum_{\rm ls}w_{\rm ls }e_{\rm RMS}^2}{\sum _{\rm ls} w_{\rm ls}}\,,
\end{equation}
can be determined with the help of variance in the ellipticity as a function of source galaxy properties \citep{Bernstein_Jarvis_2002}. This variance is provided in the shape catalog itself. Lastly, in the Eq.~\ref{eq:stacked_signal}, $\hat{m}$ represents multiplicative bias defined as $\hat{m}=\Sigma_{\rm ls}w_{\rm ls}m_s/\Sigma_{\rm ls}w_{\rm ls}$.

\subsection{Galaxy-Galaxy lensing measurements}
\label{sec:observations}
Our lens galaxy sample consists of satellite galaxies in galaxy clusters. We would like to investigate the effects of the galaxy cluster environment on the dark matter distribution around these satellite galaxies. Therefore we bin our satellite galaxies from the redMapper member galaxy catalog in four bins in cluster-centric projected comoving distances, (0.1-0.3], (0.3-0.5], (0.5-0.7], (0.7-0.9] $h^{-1}\rm Mpc$. In Table~\ref{table:satellies_data}, we list the number of satellites that fall in each of these distance bins. We calculate the weak lensing signal around these satellite galaxies in 20 logarithmically spaced comoving radial bins between $0.01-5~h^{-1}$ Mpc from the satellite galaxies, which enables us to study not only the subhalo mass of these satellites but also the halo masses of their host clusters. 

In order to understand how the cluster environment affects satellite galaxies, we also measure the weak lensing signal around a control sample of galaxies that do not reside in galaxy clusters. We construct a control sample of galaxies using the photometric galaxy catalogs from SDSS, such that they have similar $ugriz$ magnitudes as those of our satellite galaxy sample. We ensure that none of the galaxies from the control sample are members of the SDSS redMaPPer clusters or fall within the nominal cluster radius $R_{\lambda}$ provided in the redMaPPer cluster catalog \citep{Rykoff2014}. Given that the colors of galaxies depend upon the redshift of these galaxies, such a selection also helps us to select a control sample which statistically has the same redshift distribution as that of the satellite galaxy sample. For each of these satellite galaxies, we find 5 nearest neighbours in the magnitude space in order to construct a control sample of field galaxies.

\begin{figure*}
    \begin{subfigure}[b]{0.5\textwidth}
    \includegraphics[width=0.99\columnwidth]{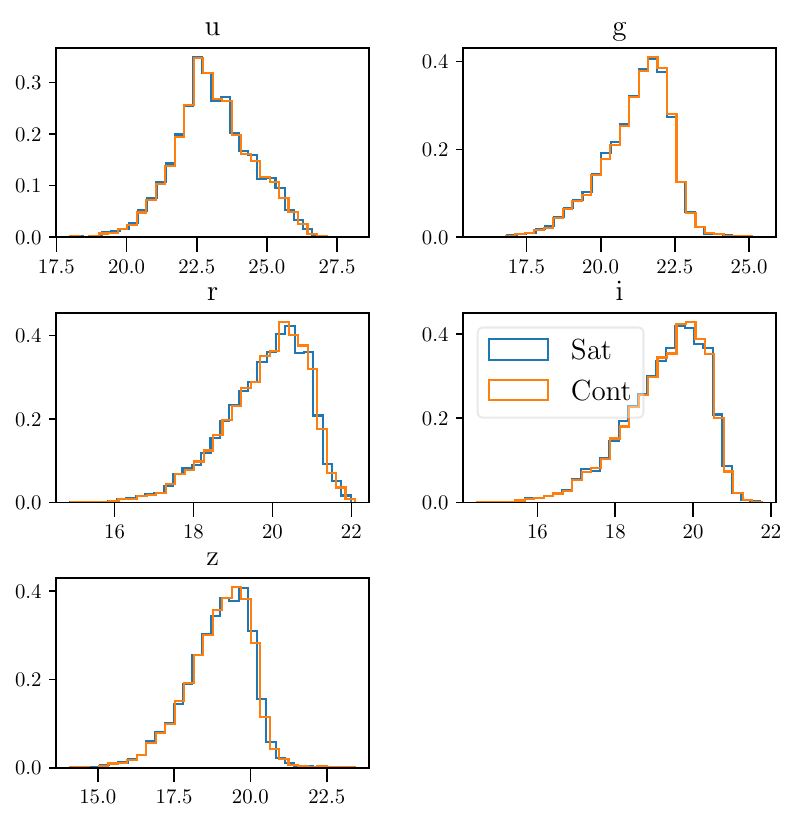}
    \end{subfigure}%
    \begin{subfigure}[b]{0.5\textwidth}
    \includegraphics[width=0.99\columnwidth]{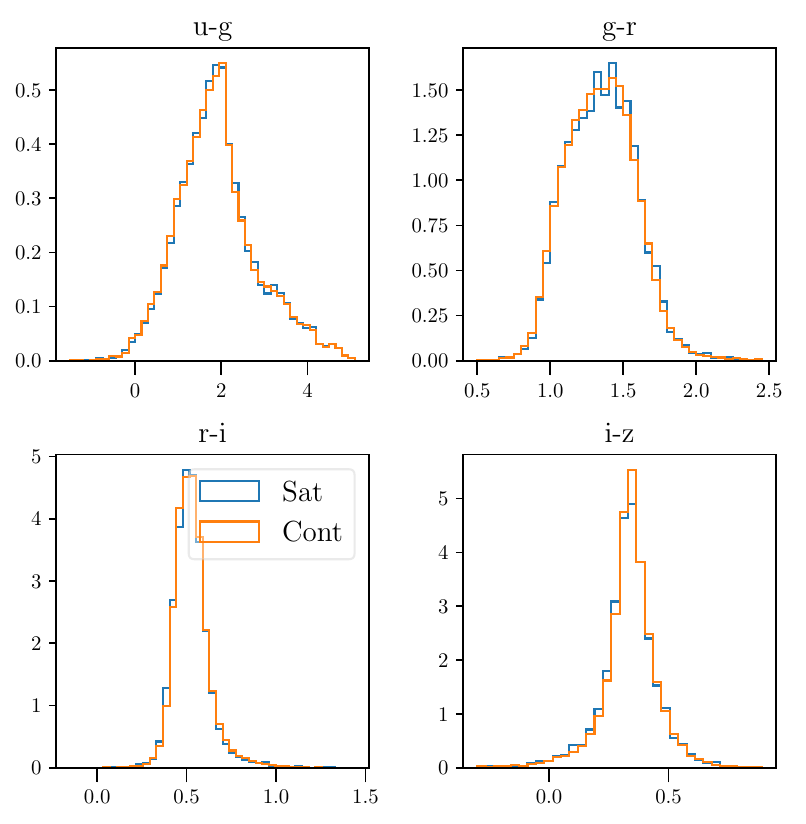}
        \end{subfigure}%
    \caption{The left hand panels show the magnitude distribution of all the satellite galaxies from redMaPPer cluster catalog (blue), and the control sample of galaxies (orange) chosen from SDSS. The right hand panels show the color distribution for both galaxy samples. The match in color and magnitude statistically allows us to match the redshift and magnitude distributions of the satellite and control sample of galaxies.}
    \label{fig:color_color_scatter_fields_satellites}
\end{figure*}

In Fig.~\ref{fig:color_color_scatter_fields_satellites}, we show the color distribution for these satellites and the control sample of galaxies in different wavelength filters to show the degree to which the colors of the control sample match with that of the satellite galaxies. Given the similarity of the control sample, we expect that these galaxies will have similar stellar mass and redshift distribution, and the differences in the weak lensing signal can be attributed to differences in the dark matter around them. We calculate the lensing signal around the control sample of galaxies in 10 logarithmic bins spaced between $0.01-0.3$ \mpch in projected comoving separation. At such distances, the contribution to the weak lensing signal is expected to be dominated by the dark matter halos of central galaxies in the control sample. 

\subsection{Weak lensing systematics}

In the left hand panel of Fig.~\ref{fig:systematics_tests}, we present the measurements of the cross component of the lensing signal around satellite galaxies as starred symbols with errors, while the round symbols show the signal around random points distributed in the same redshift range and spread over the entire sky area as our satellite galaxy sample. In the right hand panel we show similar results but corresponding to the control sample of galaxies. 

For the signal around random points corresponding to satellite galaxies, we obtain a $\chi^2$ value of $\sim 27$ for 20 degrees of freedom. This implies a p-value of $0.1
4$ to exceed $\chi^2$. The signal around random points corresponding to the control sample is $\sim 10$ for $10$ degrees of freedom, which corresponds to a p-value of $0.43$. The cross-signal around the control sample shows a $\chi^2$ value of $\sim 14$ for 10 points which corresponds to a p-value of $0.17$.

We obtain a somewhat larger value of $\chi^2\sim 32$ for the cross signal around satellite galaxies with 20 points which implies a p-value of $0.04$. This could be a result of a statistical fluctuation. The measurements suggest that the $\chi^2$ is driven by points at $R>1$ \mpch. Given that we measure distances from satellite galaxies, this could also be a result of shape measurement systematics at distances which correspond to the central regions of the galaxy clusters in which these satellites reside. Given that the galaxy cluster is a crowded environment with lots of blending, the possibility that the shape measurements in such environments could possibly be affected. As the statistical errors go down with increasing area covered by the HSC survey, we flag it as a potential systematic to be explored in the near future. Here we note that the measurements of the subhalo masses of satellite galaxies that we are interested in come from scales which are closer to the satellite galaxies and thus at radii smaller than the possibly problematic regions mentioned above and proceed with our analysis. 

We also observe that the errors in $R\Delta\Sigma_{\times}$ increase at larger radial distances. The error in this quantity is expected to be constant in size as a function of radial distance if all of the lens-source pairs are independent. However, as we are considering satellite galaxies which are spatially close to each other, the source galaxies used for the signal computation are not independent of each other, and thus the errors tend to increase as a function of distance from the satellite galaxy.

We have also computed the boost factors by computing the ratio between the number of satellite galaxy lens-source pairs and the number of random-source pairs. The boost factor can be used to correct the signals for contamination from source galaxies intrinsically at redshifts similar to the galaxy clusters, but whose photometric redshifts are much larger due to systematic and/or statistical errors. We find boost factors of the order 10 percent in the regions beyond $0.8$ \mpch (and is smaller than our relative errors), while in the inner regions we see a value for the boost factor below unity consistent with obscuration effects \citep[see e.g.,][]{Miyatake_2015}. Therefore rather than applying any boost factor correction, we have tested our signals by changing $z_{\rm diff}=0.37$, thus only using source galaxies that have their integrated probability to lie at a redshift greater than $0.7$ to be $0.99$ much more separated from lens samples. We have repeated our entire analysis and the results we obtain are consistent with the fiducial results presented in the paper given the errors.

We have computed the photometric redshift biases using a sample of COSMOS galaxies with 30 band photometric redshifts \citep{Laigle2015}, which have also been observed with HSC. These galaxies were reweighted to follow the same color and photometry difference. We follow the procedure laid out in section 3.4 in \citet{Miyatake2019}, to compute the ratio between the $\Sigma_{\rm crit}^{-1}$ estimated based on HSC photometric redshifts and that based on the better quality COSMOS redshifts \citep{Nakajima_2012}. We estimate that these biases are less than 2 percent, much smaller than the statistical uncertainty in our measurements.

 \begin{figure}
    \centering
    \includegraphics[]{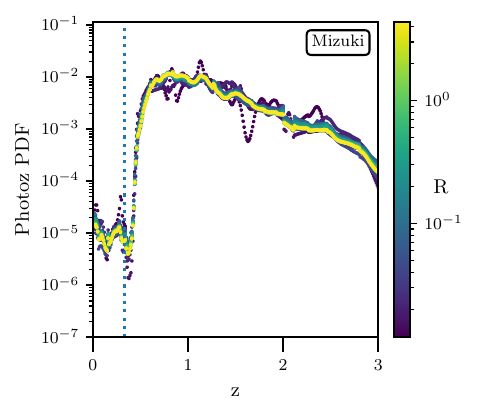}
    \caption{This figure shows the stacked photoz distribution of source galaxies in HSC field of view, around our redMaPPer satellite galaxies with separation, $R_{\rm sat}$:[0.1-0.3] from the BCG. The different colored dots represent the PDFs of sources at different radial distances away from the satellite galaxies, whereas the vertical dashed blue line marks the maximum redshift value, z$_{\rm max}$ for the lens sample we used in our analysis.}
    \label{fig:photoz_contamination}
\end{figure}

Finally, we also have computed the stacked photometric redshift posterior distribution for our weak lensing galaxy samples as a function of distance of these galaxies from the satellite lens galaxies. \citet{Varga2019} demonstrate that their weak lensing sample of DES galaxies has a substantial contribution at the redshifts of the clusters especially at small cluster-centric separations. In contrast, in Fig.\ref{fig:photoz_contamination}, we see that our weak lensing galaxy sample has negligible support at redshifts between $[0.1, 0.33]$, even when plotted with a log scale on the y-axis. This is a result of our use of a conservative cut on the photoz PDF (see Eq.\ref{eq:photoz}). The depth of HSC implies that we have a statistically large sample of galaxies, even after such a conservative cut.

\begin{table}
 \centering
 \begin{tabularx}{0.9\columnwidth} {    
  | >{\centering\arraybackslash}X 
  | >{\centering\arraybackslash}X 
  | >{\centering\arraybackslash}X | }
  \hline
  $ R_{\rm sat}[h^{-1}\rm Mpc]$& Satellite sample& Control sample\\
 \hline
 (0.1-0.3] & 949 & 4434 \\
 \hline
(0.3-0.5]& 997 & 4648 \\
 \hline
(0.5-0.7] & 977 & 4614  \\
 \hline
(0.7-0.9] & 1028 &  4821  \\[0.5ex]
\hline
\end{tabularx}
\caption{The number of redMaPPer satellite galaxies which fall in the HSC year-one footprint, have a redshift range 0.1<z<=0.33, and that are part of clusters with a BCG with P$_{\rm cen}$>0.95 are listed in each cluster-centric projected distance bin. For each of the satellite galaxies, we find 5 galaxies not residing in clusters in order to construct a control sample of galaxies with the same flux distribution in the SDSS $ugriz$ bands. The number of galaxies in the control sample are listed in the last column.}
\label{table:satellies_data}
\end{table}

 \begin{figure}
    \centering
    \includegraphics[width=0.98\columnwidth]{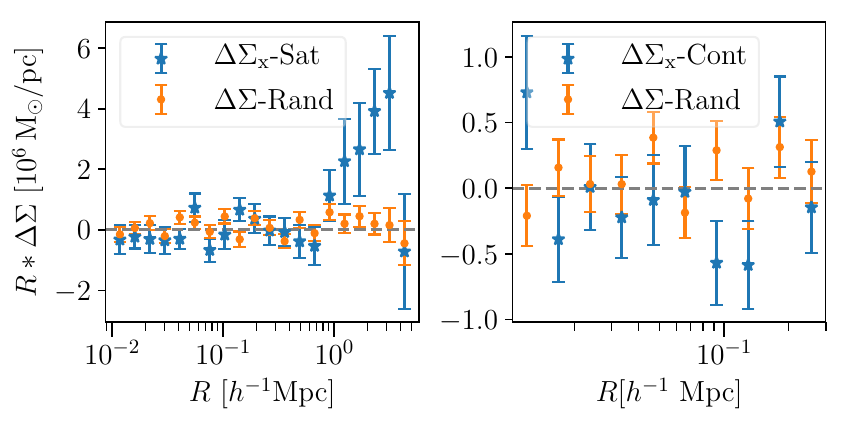}
    \caption{Systematics tests: The left hand panel shows the cross component of the lensing signal $\Delta\Sigma_{\times}$ for the satellite galaxies (shown with blue symbols with errors). The signal around random points is shown with orange symbols with error, and is also consistent with zero given the errors. The right hand panel shows similar systematics test but for the control sample of galaxies.}
    \label{fig:systematics_tests}
\end{figure}

\section{Model}
\label{sec:modelling}

In this section, we describe our model to fit the weak lensing observations for satellite and control sample galaxies described in Section~\ref{sec:observations}. 

\subsection{Halo Density Profile}
\label{sec:density_profile}
In order to describe the dark matter distribution within a halo, we use the Navarro, Frenk \& White (NFW) \citep{Navarro1997} profile given as , 
\begin{equation}
\rho(r)= \frac{\delta_{\rm c}\rho_{\rm m}}{\left(\frac{r}{r_{\rm s}}\right)\left(1+\frac{r}{r_{\rm s}}\right)^2}\,.
\label{eq:nfw}
\end{equation}
Here, $r$ represents the radial distance from the halo center, $\rho_{\rm m}= 3H^2_0\Omega_{\rm m}(1+z)^3/8\pi G$ is the mean density of the Universe at redshift z, and $r_{\rm s}$ is the scale radius of the halo. We assume that dark matter halo boundary is given by $r_{\rm 200m}$, a radius that encloses a density within the halo which is 200 times the mean matter density. Therefore the proportionality constant in the above equation $\delta_{\rm c}$ is given by, 
\begin{equation}
    \delta_{\rm c} = \frac{200}{3} \frac{c^3}{\ln(1+c)- c/(1+c)}\,.
\end{equation}
Here $c$ denotes the concentration parameter, and corresponds to the ratio between $r_{\rm 200m}$ and the scale radius $r_{\rm s}$.

We will use this model to describe the 3 dimensional density profile of dark matter belonging to the subhalos of the satellite galaxies, the control population as well as the dark matter distribution around the centers of galaxy clusters. There are indications from numerical simulations that the concentrations of subhalos get altered in response to the tidal stripping and the subsequent relaxation of the density profiles \citep{Hayashi2003, Poster_mira2011, Moline2017}. In their modelling scheme, \citet{Sifon2018} use such higher concentrations for the subhalos motivated from numerical simulations, but then quote a subhalo mass within an aperture where the density of the subhalo matches the ambient density within the main halo. Given our statistical errors, our attempts of constraining the concentrations and masses of the subhalos independently resulted in large degeneracies. Therefore, we take the simpler route of fixing the concentrations based on the masses, but an approach which can be mimicked in mocks before appropriate comparisons are made.

\subsection{Modelling lensing signal around satellite galaxies}
\label{sec:subhalo_contribution}
The weak lensing signal around a satellite galaxy in a galaxy cluster consists of the contribution from its own baryonic mass $\Delta\Sigma^{\rm bary}_{\rm sat}(R)$, the dark matter belonging to its own subhalo $\Delta \Sigma^{\rm DM}_{\rm sat} (R)$, as well as the dark matter associated with the galaxy cluster $\Delta \Sigma^{\rm DM}_{\rm clu}(R)$, whose gravitational influence dictates its motion and evolution within the halo. Therefore we can write the lensing signal around a satellite at a projected radius $R$, as a sum of three separate terms,  
\begin{equation}
\Delta \Sigma(R) =  \Delta \Sigma^{\rm DM}_{\rm sat} (R) + \Delta \Sigma^{\rm DM}_{\rm clu}(R)+ \Delta\Sigma^{\rm bary}_{\rm sat}(R)\,.
\label{eq:deltasigmatotal}
\end{equation}

\begin{figure}
    \centering
    \includegraphics[width=0.6\columnwidth]{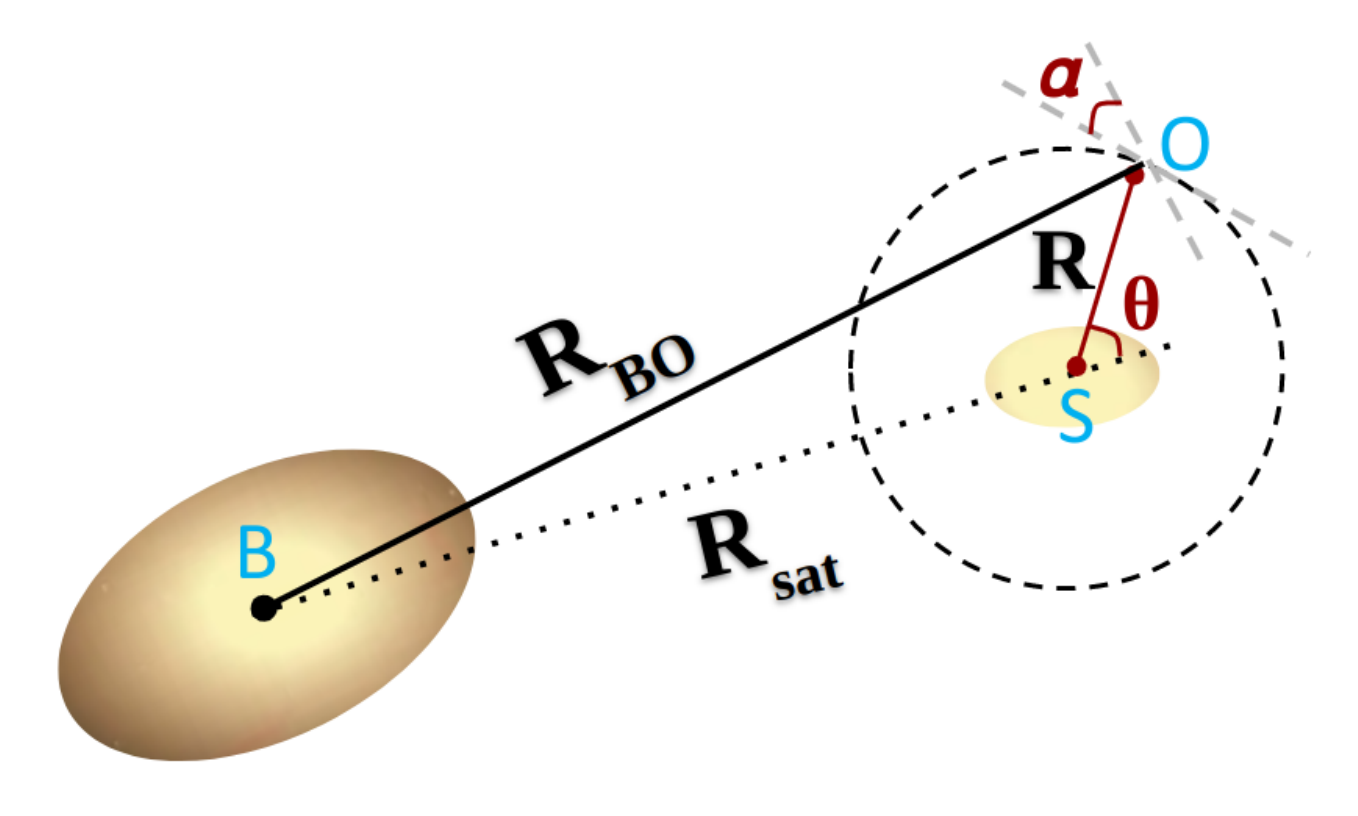}
    \caption{The figure shows the geometric configuration of a satellite galaxy in a galaxy cluster. We calculate the \esD around the satellite galaxy having separation $\rm R_{ sat}$ from the host center, in logarithmic bins of on-sky projected distance $R$, by accounting for the baryonic and dark masses of the satellite along with the dark mass of the host galaxy.} 
    \label{fig:satellite_configuration}
\end{figure}

\begin{figure}
    \centering
    \includegraphics[width=0.95\columnwidth]{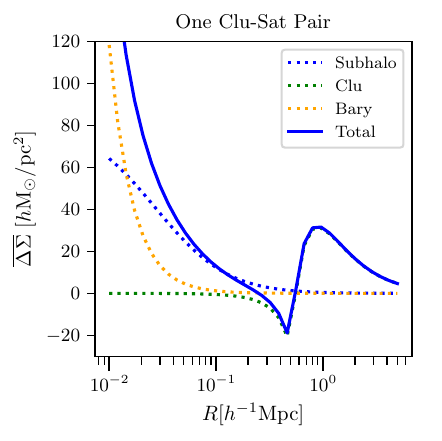}
    \caption{The figure shows the theoretical prediction for the excess surface mass density as a function of satellite centric on-sky projected distance $R$, for a satellite galaxy having separation R$_{\rm sat}$ from the BCG's center. The dashed orange (dashed-blue) line depicts the contribution due to the baryonic mass (dark matter mass) of the satellite, whereas the dashed green line shows the signal contribution from the cluster halo. The dark blue line corresponds to the overall sum of these three components. The host galaxy contributes negligibly for small values of $R$, while it becomes quite negative as $R$ approaches $\rm R_{\rm sat}$. (Fig parameters: R$_{\rm sat}$= 0.5 \mpcH, $M_{\rm clu}=10^{14.31}$, $M_{\rm sat}=10^{11.87}$, $M_{\rm bary}=10^{11.57}$ in $h^{-1}\rm M_{\odot}$)
    }
    \label{fig:individual_signal}
\end{figure}

In order to model the signal contribution due to DM mass of the satellite galaxy, i.e. the first term in Eq.~\ref{eq:deltasigmatotal}, we first obtain the projected surface density profile, $\Sigma$(R), by integrating $\rho$(r) given by Eq.~\ref{eq:nfw} along the line-of-sight,
\begin{equation}   
\Sigma_{\rm sat}^{\rm DM}(R) = \int_{-\infty}^{\infty} \rho\left(\ \sqrt{R^2+z^2} \right) dz\,.
\end{equation}
Subsequently, the average surface density within a given radius $R$ is given by,
\begin{equation}
\Sigma_{\rm sat}^{\rm DM}(<R)=  \frac{2}{R^2} \int_{0}^{R} \Sigma_{\rm sat}^{\rm DM}(R') R' dR'\,,
\label{eq:mean_sigma_daughter_inside}  
\end{equation}
and the above two equations allow us to obtain $\Delta\Sigma$ for the dark matter associated with the subhalo using Eq.~\ref{eq:esd}. We use the analytical expressions given in eq. 14 of \cite{OaxacaWright2000} to evaluate this lensing signal. In Fig.~\ref{fig:individual_signal}, we show the lensing signal as a function of the projected distance around the satellite galaxy.
 
We model the contribution from the dark matter of the galaxy cluster around the satellite galaxy as an off-centered NFW density profile. Consider a satellite ($S$) with separation $R_{\rm sat}$ from the host cluster's BCG ($B$),  as shown in the Fig.~\ref{fig:satellite_configuration}. We need to compute the average of the lensing shear due to the cluster halo at a distance $R$ away from the satellite galaxy at a location ($O$) but tangentially oriented with respect to the satellite. Under the assumption of a circularly symmetric density distribution for the cluster halo, the lensing shear is tangentially oriented with respect to the line $BO$. From geometry, we see that,
\begin{equation}
    R_{\rm BO} = \left[ R^2 + R^2_{\rm sat}+  2 R R_{\rm sat} \cos(\theta)\right]^\frac{1}{2} \,,
\end{equation}
where $R_{\rm BO}$ is the distance between $B,O$ and $\theta$ is the angle between  the line joining the cluster center and the satellite to the line joining the satellite to the position at which the lensing signal is being computed. We can transform this shear by transforming to a coordinate system oriented at an angle $\alpha$ to obtain the shear at the location $O$ but in the tangential direction with respect to the line joining $S$ and $O$. The angle $\alpha$ is given by, 
\begin{equation}
    \alpha =\arccos\left({\frac{R+R_{\rm sat} \cos(\theta)}{R_{\rm BO}}}\right)\,.
\end{equation}
Averaging over the angle $\theta$, yields the contribution of the tangential shear from the cluster halo,
\begin{equation}
 \Delta \Sigma_{\rm clu}^{\rm DM}(R)  = \frac{\displaystyle\int_{0}^{2\pi} \Delta \Sigma_{\rm clu}^{\rm NFW}(R_{\rm BO}) \cos(2\alpha)  d\theta }{2\pi}\,.
\label{eq:shear_cordinate_transform} 
\end{equation}
Here $\Delta\Sigma_{\rm clu}^{\rm NFW}(R_{\rm BO})$ is the excess surface density of a NFW halo of mass corresponding to the galaxy cluster halo at a distance $R_{\rm BO}$ from the center of this cluster which can be computed using eq.~14 from \cite{OaxacaWright2000} again, 
but by using the mass of the cluster halo.

The signal obtained for this contribution is shown with a dashed green line in Fig.~\ref{fig:individual_signal}. For values of $R<R_{\rm sat}$, the signal is nearly zero, since at these distances the difference between the surface density averaged within the radius $R$ around the satellite is not too different from the average surface density at distance $R$ from the satellite and thus the excess surface density is nearly zero. As $R\sim R_{\rm sat}$, $\Sigma_{\rm clu}^{\rm DM}(R)$ includes the increasingly larger surface density at the center of the halo which  results in a value larger than $\Sigma_{\rm clu}^{\rm DM}(<R)$, and this results in negative values of $\Delta\Sigma_{\rm clu}^{\rm DM}$. This results in a negative signal at distances close to $R_{\rm sat}$. For $R>R_{\rm sat}$, the signal starts to increase as more and more mass is enclosed within the distance $R$ and the signal reaches a positive peak at $R\sim 2R_{\rm sat}$. The signal starts to decline for $R>>R_{\rm sat}$, as the distance $R_{BO}\rightarrow R$.

A correct determination of the cluster center is important to make sure that the modeling of the cluster halo density contribution is correct. Therefore we use only those systems whose central have been assigned a probability $P_{\rm cen}>0.95$ to be central. Nevertheless, we explore how our analysis could be potentially affected by an incorrect assignment of the cluster center. We will test our conclusions against the assumptions that the true cluster center follows a Rayleigh distribution \citep{Johnston2007} around the center assigned by redMapper such that,
\begin{equation}
    P(R_{\rm off})= \frac{R_{\rm off}}{2\pi \sigma_{\rm mc}^2}\exp{\left (- \frac{R_{\rm off}^2}{2\sigma_{\rm mc}^2}\right) }\,.
\end{equation}
Here the scale parameter $\sigma_{\rm mc}$ characterizes the width of the probability distribution of the true centre around the center assigned by the redMapper algorithm. We set $\sigma_{\rm mc}$ be a free parameter in our analysis. The satellite's true cluster-centric distance, $R_{\rm sat}^{\rm true}$, is related to $R_{\rm sat}$ and $R_{\rm off}$ by the cosine law as, 
\begin{equation}
   R_{\rm sat}^{\rm true 2}=R^2_{\rm sat} + R^2_{\rm off} -2R_{\rm sat}R_{\rm off}\cos(\theta')\,. 
\end{equation}
This we use to calculate \esD contribution due to the fraction of galaxy clusters whose BCG is misidentified,
\begin{equation}\begin{split}
\Delta\Sigma'^{\rm DM}_{\rm clu}&(R|R_{\rm sat},\sigma_{\rm mc})=\\
&\quad \int_{0}^{2\pi}\int_{0}^{\infty} \Delta\Sigma^{\rm DM}_{\rm clu}(R|R^{\rm true}_{\rm sat})\; P(R_{\rm off}|{\sigma_{\rm mc}})\; dR_{\rm off}d{\theta'}\,.
\label{eq:miscentering}
\end{split}\end{equation}
Practically we carry out the integral over $R_{\rm off}$ until $10$ \mpch and we have checked that the integral is well converged by varying this value by about 30 percent.

\subsection{ Baryonic mass contribution from the satellite}
\label{sec:stellar_contribution}
In the central regions near the satellite galaxy, we also take into account the baryonic mass of the satellite galaxy and assume it to be concentrated at the center as a point mass. The $\Delta\Sigma_{\rm sat} ^{\rm bary}$($R$) for such a component drops as $R^{-2}$, as given by
\begin{equation}
    \Delta\Sigma^{\rm bary}_{\rm sat}(R)=\frac{M_{\rm bary}}{\pi R^2}\,.
	\label{eq:stellarcontribution}
\end{equation}
We depict this behaviour by the dashed orange line in Fig.~\ref{fig:individual_signal}. The baryonic mass $M_{\rm bary}$ is assumed to be a free parameter in our model.

\subsection{Average $\Delta \Sigma$ over the $R_{\rm sat}$ bin}
We will compute the weak lensing signal by stacking the signal around satellite galaxies which are located in a finite bin in cluster-centric distances. The weak lensing signal thus will be averaged over all possible satellite distances in the bin such that,     
\begin{equation}
\overline{\Delta \Sigma}(R|\sigma_{\rm mc}) = \int  P(R_{\rm sat})  \Delta \Sigma(R|R_{\rm sat},\sigma_{\rm mc})  dR_{\rm sat}\,.
\label{eq:esd_final}
\end{equation}
Here $P(R_{\rm sat})$ denotes the probability distribution of the cluster-centric distances of the satellite galaxies in a given bin, and $\Delta\Sigma$ is computed as,
\begin{equation}\begin{split}
\Delta \Sigma(R|R_{\rm sat},\sigma_{\rm mc})  =  \Delta \Sigma^{\rm DM}_{\rm sat}(R) +  \Delta \Sigma'^{\rm DM}_{\rm clu}&(R|R_{\rm sat},\sigma_{\rm mc})\\ & +\Delta\Sigma^{\rm bary}_{\rm sat}(R)\,.
\end{split}\end{equation}
The feature seen at the cluster-satellite distance in the weak lensing signal around a single satellite galaxy gets smeared out as we average the signal around many satellites in a given cluster-centric distance bin. Similarly the bump seen at $\sim2R_{\rm sat}$ also gets broader.

\subsection{Modelling signal around the Control Sample of galaxies}
\label{sec:field_model}
In order to quantify the environmental effects on the DM distribution around satellite galaxies in galaxy clusters, we need to compare the weak lensing signal with that around a control sample of galaxies that follows the same mass and redshift distribution, but do not reside in galaxy clusters. We expect these to evolve differently than satellite galaxies in clusters, as processes specific to the dense environment are not expected to play any role in the their evolution. While modelling the control sample of galaxies, we don't need to account for contribution due to  $\Delta \Sigma_{\rm clu}^{\rm DM}(R)$, i.e. the second term in Eq.~\ref{eq:deltasigmatotal}. Hence we can re-write Eq.~\ref{eq:deltasigmatotal} for the control sample of galaxies as, 
\begin{equation}
\begin{split}
\Delta \Sigma_{\rm cont}(R) &=  \Delta \Sigma_{\rm cont}^{\rm DM}(R) +  \Delta\Sigma^{\rm bary}_{\rm cont}(R)\,.
\label{eq:esd_fields}
\end{split}
\end{equation}
We include separate parameters to describe the halo mass and the baryonic component of the control sample of galaxies. We assume that the distribution of the matter around these galaxies also follows the NFW profile as in Section~\ref{sec:subhalo_contribution} and that the baryonic component is assumed to be a point mass contribution as in \ref{sec:stellar_contribution}. If our control sample of galaxies has the same baryonic mass as the satellite galaxies in clusters, we expect the weak lensing signal in the innermost regions ($R< 0.03$ \mpch) to be similar to that of the satellite galaxies. Any difference in the signals seen on intermediate scales ($R \in [0.03, 0.1]$ \mpch), where the halo (subhalo) of the control (satellite) galaxies dominate, can be used to probe the environmental effects on the dark matter content around satellite galaxies.

 \section{Statistical Error Estimates and model fits}
 \label{sec:stastical_estimates_and_model_fits}
 \subsection{Statistical Error Estimates}
 \label{sec:Statistical_Error_Estimates}

\begin{figure*}
        \begin{subfigure}[b]{0.5\textwidth}
            \includegraphics[width=0.95\columnwidth]{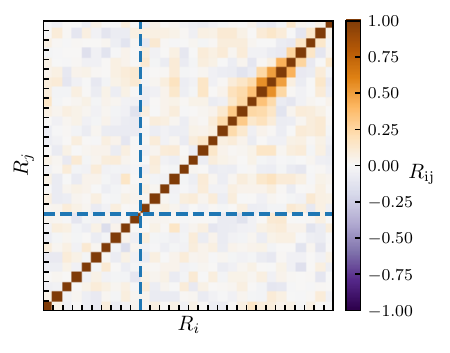}
        \end{subfigure}%
        \begin{subfigure}[b]{0.5\textwidth}
            \includegraphics[width=0.95\columnwidth]{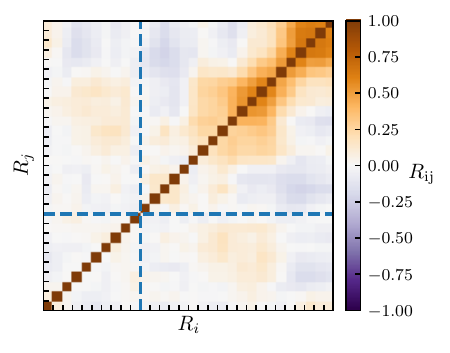}
        \end{subfigure}%
        \caption{{In this figure we show the joint covariance matrices obtained for the satellites and their counterpart control sample galaxies, for one of the radial bins, i.e. 0.1-0.3 $h^{-1} \rm Mpc$ used in our analysis, using (i) random rotations to capture the shape noise (left panel) and (ii) Jackknife (right panel). Other radial bins show similar behaviour. There is little cross-covariance between the signal around satellites and the control sample. Therefore, the covariance matrix has a block diagonal form where individual blocks correspond to the covariance matrices for the satellites and control sample, respectively.}}
    \label{fig:covariance_matrices}
\end{figure*} 

The weak lensing signal manifests itself as a coherent shear on the shapes of galaxies. The dominant statistical uncertainty in the measurement of the weak lensing signal on small scales is due to shape noise, the presence of intrinsic ellipticity in the shapes of the finite number of source galaxies (from HSC) that lie behind the lens galaxies (in our case, control sample and satellite galaxies). The shape noise term scales as ${N^{-1/2}}$ where $N$ is the number of pairs of lens-source galaxies. Furthermore, a given source galaxy can lie behind multiple lens galaxies at different distances. This can lead to covariance between the weak lensing signal measurements on different scales. We obtain an estimate of the shape noise covariance by performing $320$ random rotations of the shapes of source galaxies. We compute the lensing signal around lens galaxies in each of these realizations and use its dispersion to compute the shape noise covariance. 

We also use the jackknife technique \citep{Miller1974} to estimate this covariance by dividing the redMapPer random catalog which overlaps with the HSC survey area into 88 approximately equal sized regions. We then compute the signal by leaving each region out, and obtain the jackknife covariance. We have confirmed that for each of our bins the covariance derived from the random rotation technique and from the jackknife agrees well on small scales, while on large scales the jackknife technique leads to a larger covariance. This is expected as the jackknife technique is able to capture the covariance due to large scale structure.  In this paper, we use the jackknife covariance for model fits. The cross-correlation matrix is given by
\begin{equation}
    {\cal R}_{ij} = \frac{C_{ij}}{[C_{ii}C_{jj}]^{1/2}}\,,
    \label{eq:correlation_cofficients}
\end{equation}
and it captures the extent of the covariance between different bins. We show the cross-correlation matrices corresponding to the shape-noise and jackknife technique for one of the used radial bin i.e. 0.1-0.3 $h^{-1} \rm Mpc$, in Fig.~\ref{fig:covariance_matrices}. Given that we compute the lensing signal, \esD  in 20 logarithmic spaced radial bins in the range $ [0.01, 5]\ \rm h^{-1}Mpc$, the size of the errors at large values of R, decreases significantly as the number of lens-source pairs in those annuli increases substantially due to the proportionate increase in area. To smooth out noise in the Jackknife covariance matrix, we smooth the cross-correlation matrix using a box-car filter of size 3 following the procedure adopted in \cite{mandelbaum2013}.

\begin{table}
\renewcommand{\arraystretch}{1.5}
 \centering
 \begin{tabularx}{0.9\columnwidth} {    %fixing width of the table
  | >{\centering\arraybackslash}X 
   |>{\centering\arraybackslash}X | }
  \hline
   \multicolumn{2}{c}{Model Parameters}\\[0.5ex]
  \hline
  Parameters&Priors\\
  \hline
  $\rm \log[M_{\rm clu}/(h^{-1}M_{\odot})]$ & flat[10,16] \\
  $\rm \log[M_{\rm sat}/(h^{-1}M_{\odot})]$& flat[9,16]\\
  $\rm \log[M_{\rm cont}/(h^{-1}M_{\odot})]$&flat[9,16]\\
  $\rm \log[M_{\rm bary}/(h^{-1}M_{\odot})]$&flat[8,14]\\
  $\rm \sigma_{\rm mc}[h^{-1} {\rm Mpc}]$& flat(0,1)\\
  $\rm f_{\rm orp}$ & flat[0,1]\\[0.5ex]

\hline
\hline
\end{tabularx}
\caption{The table shows the uninformative priors that we use for sampling the posterior distribution of the model parameters in the MCMC analysis. The parameters $M_{\rm clu},M_{\rm sat}, M_{\rm cont}$ denote the halo (subhalo) masses of the main cluster, satellite galaxies and the control sample of galaxies, respectively. The parameters $\rm M_{\rm bary}$ denotes the average baryonic mass of the satellite as well as that of the control sample of galaxies. Finally, $\rm \sigma_{\rm mc}$, is the parameter which quantifies the width of the 2-D Rayleigh probability distribution used for the mis-centering analysis, while the parameter $\rm f_{\rm orp}$ represents the fraction of orphan satellites. }

\label{table:priors}
\end{table}

\subsection{Model fits}
\label{sec:fitting_procedure}
We compute the lensing signal using Eq.~\ref{eq:esd} around satellite galaxies located in four cluster centric projected distance bins as described  in Section~\ref{sec:observations}. We construct a physically motivated model for the satellites and the control sample, which we described in Section~\ref{sec:modelling}. The signal around the control sample is modelled with only 2 parameters -- the mass of its dark matter halo plus a baryonic mass component. Our model for the signal around the satellite galaxies comprises of 4 parameters, i.e. the mass of the satellite subhalo, the baryonic mass of the satellite galaxy, the mass of the main cluster halo as well as a parameter $\sigma_{\rm mc}$, which characterizes the width of the Rayleigh probability distribution used for accounting the  miscentering (Eq.~\ref{eq:miscentering}) of the central galaxy in our clusters.

We use the concentration-mass relation from \cite{Macci_2007} encoded in the publicly available repository AUM \footnotemark \footnotetext{https://github.com/surhudm/aum}, to get the concentration from the mass of the halo. However, we note that even including the concentration as a free parameter do not impact our constraints on the masses. We discuss the flat prior ranges used for all our parameters in the Table:\ref{table:priors}.

Having specified the model for the observed weak lensing signals, we obtain the posterior distribution of the parameters ($\Omega$) given the observed data vector (${\bf D}$) as,
\begin{equation}
    P(\Omega|{\bf D}) \propto P({\bf D}|\Omega) P(\Omega)\,.
    \label{eq:posterior}
\end{equation}
Here $P({\bf D}|\Omega)$ is the likelihood and $P(\Omega)$ is the prior probability distribution of the model parameters. We assume the likelihood to be a Gaussian such that, 
\begin{equation}
    \centering
    \ln P({\bf D}|\Omega) = -\frac{1}{2} \chi^2 = -\frac{1}{2}{(\textbf{D}-\bar{\bf D})^{\textbf{T}} C_{ij}^{-1} (\textbf{D}-\bar{\bf D})}\,.
    \label{eq:likelihood}
\end{equation}
Here $\bar{\bf D}$ corresponds to the vector of model predictions, and we use the smoothed covariance matrix, $C_{ij}$ obtained with the jackknife technique as described in Section~\ref{sec:Statistical_Error_Estimates}.

We perform a Monte Carlo Markov Chain(MCMC) analysis with the help of publicly available emcee\footnotemark \footnotetext{https://github.com/dfm/emcee}
package \citep{mcmc_hogg_2013} in order to sample from the posterior distribution of our model parameters. For the MCMC analysis, we use uninformative flat priors (see Table~\ref{table:priors}), which span a wide sample space, in order to obtain data driven constraints on the parameters. We use 100 walkers running over 20000 steps, along with  500 burn-in steps which we discard, to allow the chain to reach an equilibrium state.

\section{Results and Discussion}
\label{sec:results_and_discussion}
\begin{figure*}
  \includegraphics[width=0.99\textwidth]{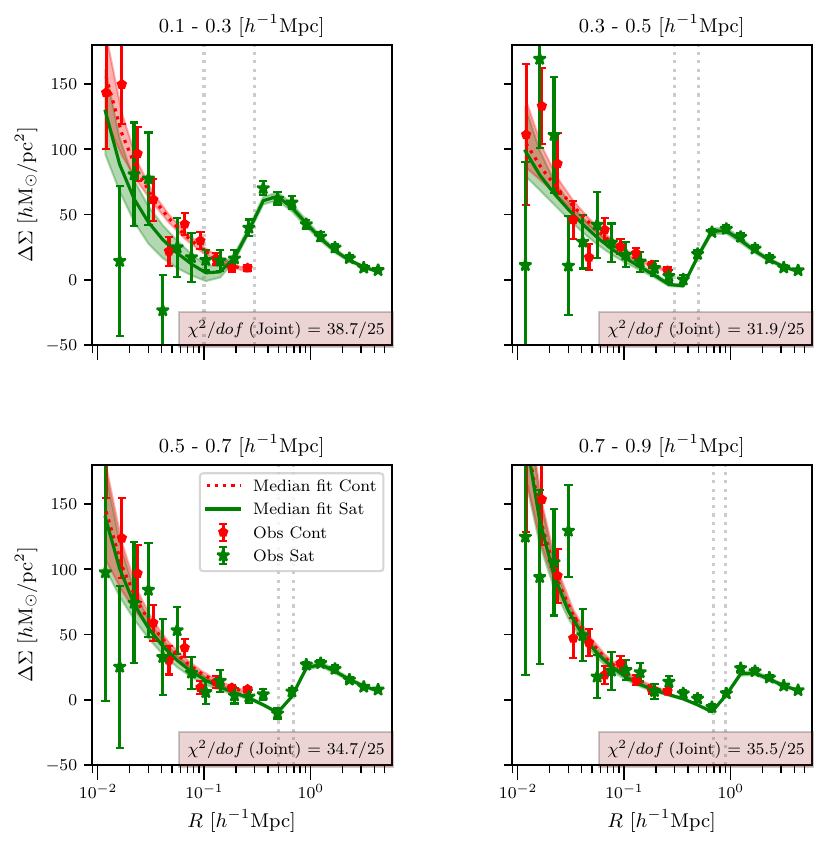}
  \caption{The figure shows the galaxy-galaxy lensing signal around satellite galaxies (green symbols with errors) and the control sample of galaxies (red symbols with errors) in different radial bins of R$_{\rm sat}$, as indicated by the title of each subpanel. The green (red) shaded regions show the 68 percent credible region around the median, using the parameters obtained from MCMC.  The dotted grey lines represent the cluster centeric distance limits for each selection bin as mentioned in the title. The inferred parameter constraints are listed in the Table~\ref{table:inferred_parameters}.}%\textbf{ This dof is not the Raveri and Hu 2018, is it okay? }} 
\label{fig:signal_all_bins}
\end{figure*}

\begin{figure*}
        \begin{subfigure}[b]{0.5\textwidth}
        \includegraphics[width=0.99\columnwidth]{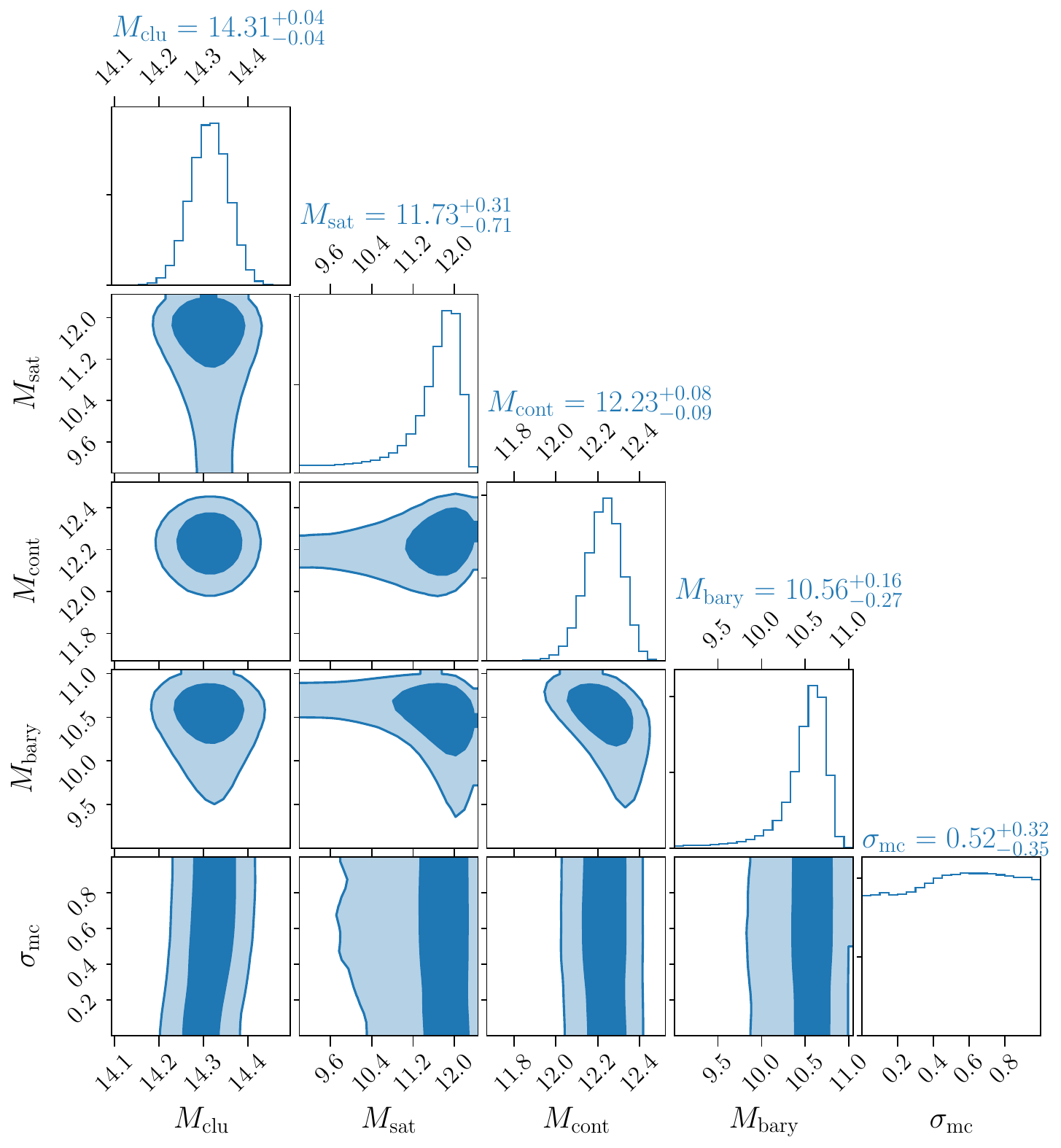}         
        \end{subfigure}%
        \begin{subfigure}[b]{0.5\textwidth}
            \includegraphics[width=0.99\columnwidth]{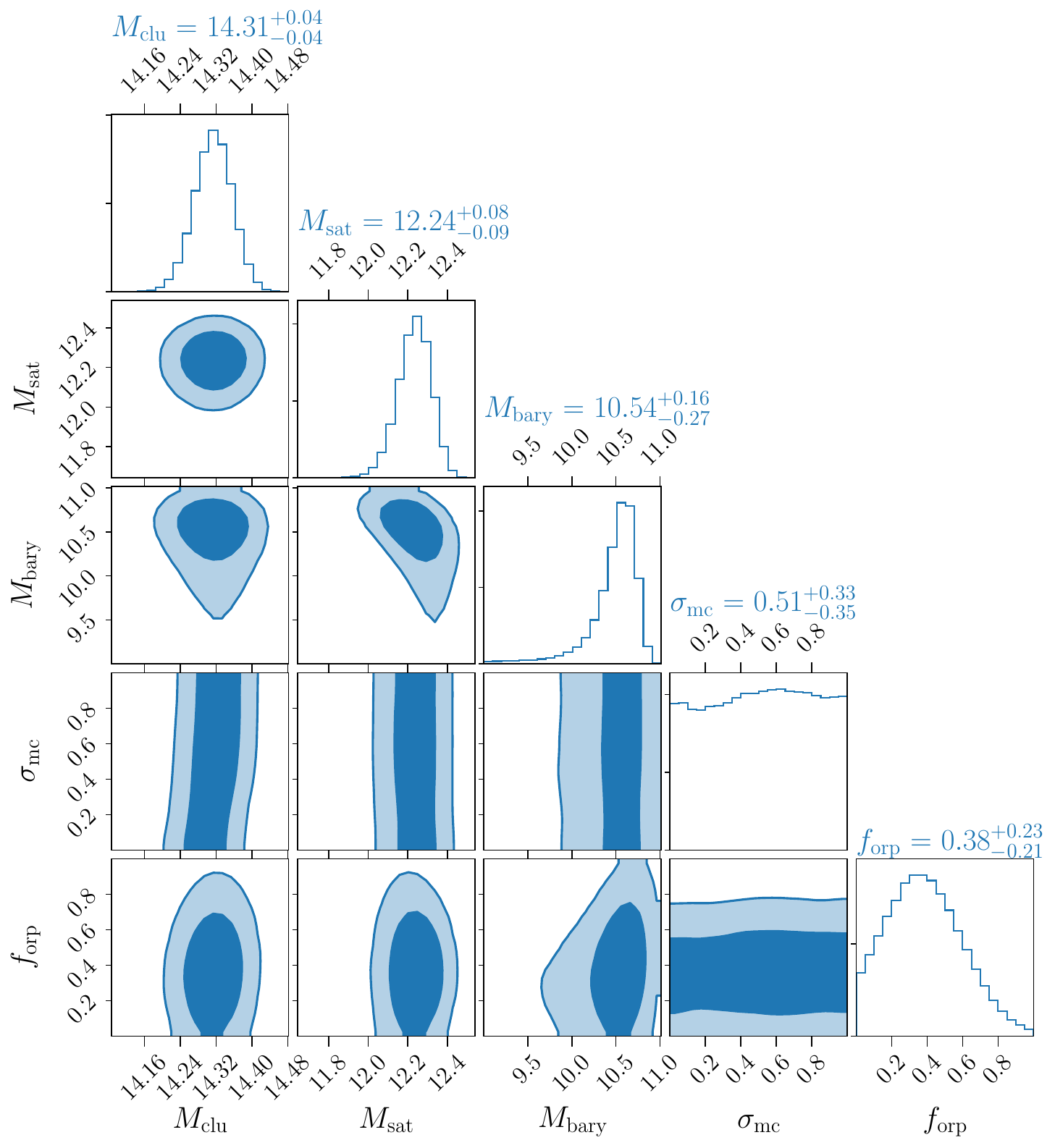}
        \end{subfigure}
        \caption{ The figure shows the posterior distribution of various model parameters used to joint fit the lensing signal  around the satellites and their counterpart control sample of galaxies, having separation of 0.1-0.3 $h^{-1} \rm  Mpc$ from cluster center. The corner plots in the left and right panels represents the parameters for the analysis of subhalo mass estimation (refer Section~\ref{sec:mass_disruption}) and  orphan fraction (refer Section~\ref{sec:orphan_fraction}) respectively. In the figure the contours represents 68 and 95 credible intervals and the median values with their 1$\sigma$ are given for each parameter at the top of the posteriors. The figure is representative of what we observe in other bins too, and summarised in Table: \ref{table:inferred_parameters}}
        \label{fig:Posterior_distribution}
\end{figure*}

\begin{table*}
\centering 
\renewcommand{\arraystretch}{1.5}
\rule{0pt}{4ex}
\begin{tabularx}{\textwidth} {    %fixing width of the table
   >{\centering\arraybackslash}X 
   >{\centering\arraybackslash}X 
   >{\centering\arraybackslash}X 
   >{\centering\arraybackslash}X 
   >{\centering\arraybackslash}X 
   >{\centering\arraybackslash}X 
   >{\centering\arraybackslash}X 
   >{\centering\arraybackslash}X}
  \hline 
  $\rm R_{sat} ~$[$h^{-1}\rm Mpc$]& $ \rm \log \dfrac{M_{\rm clu}}{[~h^{-1}M_{\odot}]}$ & $ \rm \log \dfrac{M_{\rm sat}}{[~h^{-1}M_{\odot}]}$ & $ \rm \log \dfrac{M_{\rm cont}}{[~h^{-1}M_{\odot}]}$ & $ \rm \log \dfrac{M_{\rm bary}}{[~h^{-1}M_{\odot}]}$ &  $\sigma_{\rm mc} [h^{-1}\rm Mpc]$& $ \rm \log \dfrac{M^{Mizuki}_{\rm stel}}{[~h^{-1}M_{\odot}]}$ \Tstrut\Bstrut\\
 \hline\hline
 (0.1-0.3] & $14.31^{+0.04}_{-0.04}$ & $11.73^{+0.31}_{-0.71}$ & $12.23^{+0.08}_{-0.09}$ & $10.56^{+0.16}_{-0.27}$ &  $0.52^{+0.32}_{-0.35   }$&  $10.48^{+0.31}_{-0.31 }$\\
 (0.3-0.5] & $14.33^{+0.04}_{-0.04}$ & $12.11^{+0.10}_{-0.14}$ & $12.24^{+0.06}_{-0.06}$ & $10.07^{+0.37}_{-0.63}$ & $0.47^{+0.34}_{-0.28}$ & $  10.46^{+0.32}_{-0.32 }$ \\
(0.5-0.7]  & $14.36^{+0.04}_{-0.04}$ & $12.00^{+0.12}_{-0.17}$ & $12.11^{+0.08}_{-0.09}$ & $10.52^{+0.18}_{-0.34}$ & $0.44^{+0.37}_{-0.30}$ &$  10.50^{+0.31}_{-0.31 }$\\
(0.7-0.9]  & $14.39^{+0.04}_{-0.06}$ & $12.07^{+0.09}_{-0.11}$ & $12.14^{+0.07}_{-0.08}$ & $10.77^{+0.10}_{-0.13}$ & $0.40^{+0.38}_{-0.29}$ & $  10.51^{+0.30}_{-0.30 }$\\[0.5ex]
\hline
\end{tabularx}
\rule{0pt}{0.01ex}
\begin{tabularx}{\textwidth} {    %fixing width of the table
   >{\centering\arraybackslash}X 
   >{\centering\arraybackslash}X 
   >{\centering\arraybackslash}X 
   >{\centering\arraybackslash}X 
   >{\centering\arraybackslash}X 
   >{\centering\arraybackslash}X 
   >{\centering\arraybackslash}X 
   >{\centering\arraybackslash}X}
  \hline 
  $\rm R_{sat} ~$[$h^{-1}\rm Mpc$]& $ \rm \log \dfrac{M_{\rm clu}}{[~h^{-1}M_{\odot}]}$ & $ \rm \log \dfrac{M_{\rm sat}}{[~h^{-1}M_{\odot}]}$ & $ \rm \log \dfrac{M_{\rm bary}}{[~h^{-1}M_{\odot}]}$ & $\sigma_{\rm mc} [h^{-1}\rm Mpc]$ &$f_{\rm orp}$& $ \rm \log \dfrac{M^{Mizuki}_{\rm stel}}{[~h^{-1}M_{\odot}]}$\Tstrut\Bstrut\\
\hline\hline
 (0.1-0.3] & $14.31^{+0.04}_{-0.04}$ & $12.24^{+0.08}_{-0.09}$ & $10.54^{+0.16}_{-0.27}$ &  $0.51^{+0.33}_{-0.35}$ & 0.48 &  $10.48^{+0.31}_{-0.31 }$\\
 (0.3-0.5] & $14.32^{+0.04}_{-0.04}$ &  $12.24^{+0.06}_{-0.06}$ &  $10.09^{+0.36}_{-0.64}$ & $0.47^{+0.34}_{-0.28}$ & 0.22  & $  10.46^{+0.32}_{-0.32 }$\\
(0.5-0.7]  & $14.36^{+0.04}_{-0.04}$ & $12.11^{+0.08}_{-0.09}$ & $10.54^{+0.17}_{-0.33}$ & $0.44^{+0.37}_{-0.30}$  & 0.20 &$  10.50^{+0.31}_{-0.31 }$\\
(0.7-0.9]  & $14.39^{+0.04}_{-0.05}$ & $12.14^{+0.07}_{-0.08}$ & $10.78^{+0.10}_{-0.13}$ & $0.40^{+0.38}_{-0.29}$ & 0.12 & $  10.51^{+0.30}_{-0.30 }$\\[0.5ex]
\hline
\end{tabularx}

\caption{The table shows the posterior distribution of the parameters characterized by their median values along with errors (based on 16 and 84 percentile), used in analysis of subhalo mass estimation (refer Section~\ref{sec:mass_disruption}), and  orphan fraction (refer Section~\ref{sec:orphan_fraction}) respectively. In the case of the orphan fraction parameter, we show the upper limit by reporting the 68 percentile value. The last column of the table shows the stellar mass estimates for the satellite galaxy samples based on the template fits of the Mizuki photometric redshift estimation code. Note that the errors here reflect the scatter in stellar masses rather than the error on the mean. Stellar masses typically have units of $h^{-2}M_\odot$, but the weak lensing signal is sensitive to $h^{-1}M_\odot$. We have converted the Mizuki stellar mass estimates with the appropriate $h$ factors to account for this difference.} 
\label{table:inferred_parameters}
\end{table*}

\subsection{Weak lensing subhalo masses of satellite galaxies}
\label{sec:mass_disruption}

In the various panels of Fig.~\ref{fig:signal_all_bins}, we show the weak lensing signal measured around the satellite galaxies at different cluster-centric distances using green symbols with errors. The signal around the control sample of galaxies is shown with red symbols. We also subtract the signal around random points, both from satellites and the control sample to account for any residual large scale systematics in the shape measurements. The signal-to-noise ratio (SNR) of our measurements around the satellite galaxies in the radial range $r\in[0.02, 0.3]$ \mpch\footnote{The subhalo mass contribution dominates the weak lensing signal in the radial range we consider to quote the signal-to-noise ratio.} is 7.39, 5.09, 4.32 and 7.53 in each of the cluster-centric distance bins, respectively.  For the corresponding control sample of galaxies in each of the bins,  the SNR in the same radial range is 11.08, 13.00, 11.14 and 9.93, respectively. As expected from our discussion in Section~\ref{sec:subhalo_contribution}, we see a distinct dip in the weak lensing signal around satellite galaxies corresponding to each cluster centric radial distance bin (shown by vertical dashed lines).

Qualitatively, we also see that for satellites which lie at large cluster-centric distances, the weak lensing signal approaches that of the control sample of galaxies. However, at distances closer to the centers of galaxy clusters, interesting differences start to emerge, especially in the first cluster-centric distance bin. The signal around the control sample appears, although noisy, appears to be systematically larger than that around satellite galaxies.

To quantify this difference we fit the model described in Sec~\ref{sec:modelling}, to the weak lensing signal measured around satellite and the control sample of galaxies, where each of these have a free parameter for the mass of their individual halos (subhalos in the case of satellite galaxies). Since our control sample is well matched to the satellites (see Fig.~\ref{fig:color_color_scatter_fields_satellites}), we use the same value of baryonic mass ( i.e. M$_{\rm bary}$ parameter in MCMC sampling) to fit for the baryonic mass of both the samples. 

We construct  joint data vectors \textbf{D} and $\bar{\bf D}$, used in the equation \ref{eq:likelihood}, with combined information from the the satellites and the control sample.

The median fit  model along with its 68 percent confidence regions from MCMC sampling, for both the samples in each case is shown with the dotted red line for the control sample and the green line for the satellite galaxy sample. Our model is able to qualitatively and quantitatively capture the various features seen in the weak lensing signal. Although the reduced $\chi^2$ we obtain in some of the cases is little high, this is attributable to the large scales where the signal from the galaxy cluster halo dominates and thus does not cause any appreciable effect on our inferences and conclusions. We note that nevertheless the cluster halo masses that we obtain are comparable to measurements in the literature \citep{Simet2017, Murata2018}, even though we effectively are restricting ourselves to large scales.

We compare the model parameter constraints on the halo/subhalo masses of satellite galaxies to that of the control sample in Table~\ref{table:inferred_parameters}. The various subpanels in the left hand side of Fig.~\ref{fig:Posterior_distribution}, show the posterior distributions of our model parameters with 68 and 95 confidence contours for the closest cluster-centric radial bin. The masses of the subhalos and the halos of the control sample are determined very precisely. In particular, we see that there is very little degeneracy between the mass of the main cluster halo and that of the mass of the subhalo of the satellite galaxy. We find the expected degeneracy between the baryonic mass of the galaxies and their subhalos. Importantly the nuisance parameter corresponding to miscentering does not show any degeneracy with the halo and subhalo masses of the control and satellite sample of galaxies.

\begin{figure}
   \centering
  \includegraphics[scale=0.5]{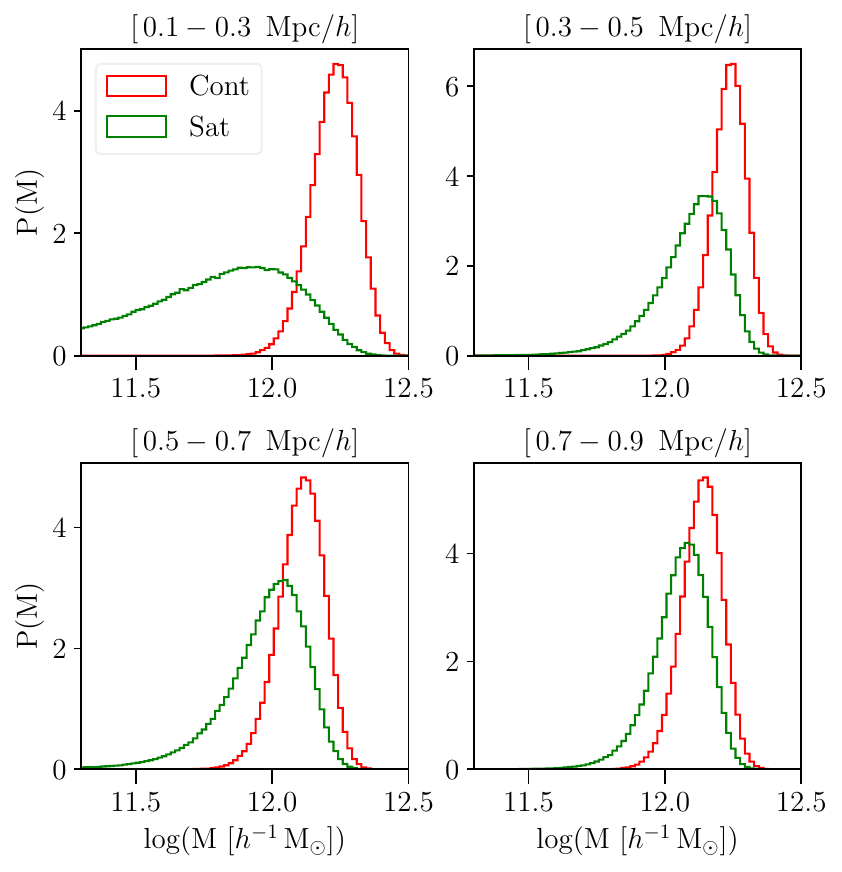}
\caption{In this figure we show the probability density of subhalo/halo mass distribution for the satellites(green) and the control sample galaxies(red), as obtained from our MCMC analysis, in different cluster centric radial bins as indicated by the title of the each sub-panel. We observe that the halo masses for the control sample exceed from the subhalo masses of the satellites with as large as factor two, specially for small distances from the cluster center.}
\label{fig:mass_histogram}
\end{figure}

In Fig.~\ref{fig:mass_histogram}, we show the posterior distributions of the masses of the subhalos of the satellite galaxies and the halos of control sample of galaxies. We see hints of differences as we move closer to the cluster center, with the most visible difference seen in the cluster centric distance of $(0.1, 0.3]$ \mpch. In this bin, the halos of the control sample are approximately a factor two larger than that of the satellite galaxies, albeit with large errors. These differences progressively reduce as we move further and further away from the center. Our results support a reduction in the dark matter masses of the satellite galaxies within galaxy clusters that depends upon the cluster-centric distance. 

\begin{figure}
  \centering
  \includegraphics[scale=0.99]{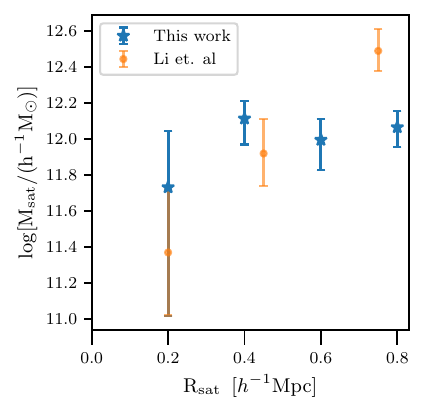}
\caption{This figure shows constraints on the logarithmic subhalo masses of the satellite galaxies with errors as blue symbols for different radial bins used in our analysis. As a comparison, we also show constraints from \citet{Li2016} from the CS82 survey for redMaPPer satellite galaxies with orange symbols. The results from both studies are consistent with each other within errors, especially in the inner regions. Small differences observed in the outer regions are likely due to subtle differences in sample selection used in both the studies.}
\label{fig:mass_as_Rsat}
\end{figure}

The dependence of subhalo masses of satellite galaxies on their cluster-centric distance can arise from multiple physical processes. Dynamical friction is expected to bring the most massive satellites to the inner regions of the halo, while tidal effects can strip the dark matter subhalos of the matter in their outskirts. In Fig.~\ref{fig:mass_as_Rsat}, we present the subhalo mass constraints as a function of cluster centric radius from our analysis as blue symbols with errors. We do not see a large dependence of the subhalo mass of redMaPPer satellites on their cluster-centric distance. We compare these results with those obtained by \cite{Li2016} using the CFHT Stripe 82 survey (orange symbols with errors). In the inner regions, the mass constraints from both the studies are consistent with each other, given the errors. The subhalo masses inferred for satellites from our measurements in the outermost bin are smaller compared to \citet{Li2016}. There are subtle sample selection differences in the redMaPPer satellite galaxies used in these analyses, e.g., \citet{Li2016} use all clusters irrespective of their $P_{\rm cen}$ values and model the miscentering. They also select clusters in a larger redshift range $z<0.5$. The selection of clusters above $z>0.33$ results in satellites below the flux limit of SDSS to be missed in the photometric catalogs, and could be a potential source for the differences we observe.

\citet{Sifon2018} analyzed the weak lensing signal around satellite galaxies within galaxy clusters from the MENeaCS \citep{Sand2012} using imaging data from the Canada France Hawaii Telescope. The stellar masses of the redMaPPer satellite galaxies should on average correspond to their third and fourth bins in stellar mass. The posterior distribution of the masses of the subhalos they obtain $\log M \sim 11.6^{+0.15}_{-0.15}\  h^{-1}M_{\odot}$ are broadly consistent, albeit $\sim 2-\sigma$ lower compared to our estimates of the subhalo masses averaged over all distances within the halo. This could possibly be due to the $\sim 10$ times more massive cluster sample they use, providing a denser environment compared to the redMaPPer cluster sample we use. They also observe a similar trend for the subhalo masses of their satellite galaxies as a function of the distance to the cluster centre as we do, where the subhalos closest to the center have roughly $3\sigma$ smaller masses.

In Table~\ref{table:inferred_parameters}, we also present the average and the standard deviation of the stellar mass estimates for the satellite galaxy samples based on the template fits of the Mizuki photometric redshift estimation code. These stellar mass estimates are broadly consistent with our determinations from weak lensing. We also note that we expect some degeneracies between our measurements of the baryonic mass and the subhalo mass, assumed concentration parameters, and any deviations of the subhalo profile from the assumed NFW profile in our analysis, especially in the inner regions.

\subsection{Upper limits on the fraction of orphan satellite galaxies}
\label{sec:orphan_fraction}
The dark matter halos of satellite galaxies get disrupted due to stripping, especially if the pericenters of their orbits bring them closer to the cluster center. Some fraction of the satellite galaxies in numerical simulations could lose almost all of their dark matter halos during such close pericentric passages, many times due to limited mass and force resolution \citep{vandenBosch2017, vandenBosch2018, vandenBosch2018b}. Such galaxies which have lost a significant portion of their dark matter halos are called orphan galaxies. Inclusion or exclusion of orphan galaxies can affect the inferred relations between stellar mass and the peak maximum circular velocities in studies which model the abundance of galaxies and their clustering \citep{Behroozi_2019}.

In this section, we use our measurements to put an observational upper limit on the fraction of such orphan galaxies directly from weak lensing observations. We assume that a fraction of satellite galaxies, $f_{\rm orp}$, have lost all their dark matter halos around them, while the rest behave similar to the control galaxy population. This is an undoubtedly simplistic model which is clearly not physical, but it allows us to put an upper limit on the fraction of satellite galaxies that could have lost their dark matter halos, which can act as a useful empirical constraint on models which invoke orphan galaxies. Studies relying on the orphan populations of galaxies, should not exceed the empirical fractions as a function of cluster-centric distances constrained in this paper.

To explore this limiting case, we perform a joint analysis to fit the signal around the satellites and the control sample galaxies simultaneously. We deliberately set the parameters corresponding to the dark matter halos of the satellite and the control sample galaxies to be the same, but include an extra parameter $f_{\rm orp}$ that describes the fraction of satellite galaxies which are orphans. We use an uninformative prior for $f_{\rm orp}$ between $[0, 1]$ in order to perform this joint analysis. We list the best fit parameters along with their 68 percent credible intervals in Table~\ref{table:inferred_parameters}. In the case of the orphan fraction parameter, we report the upper limit by calculating the 68 percentile. In Fig.~\ref{fig:orphan_fraction}, we present the constraints on the upper-bound of the orphan satellite fraction in different radial bins as blue symbols with errors. As expected, we find that the upper limits on the fraction of orphan satellites decreases with distance from the cluster center, and is the first direct empirical constraint using weak lensing observations so far.

\subsection{Effects of contamination of satellite sample}
\begin{figure}
    \centering
    \includegraphics[]{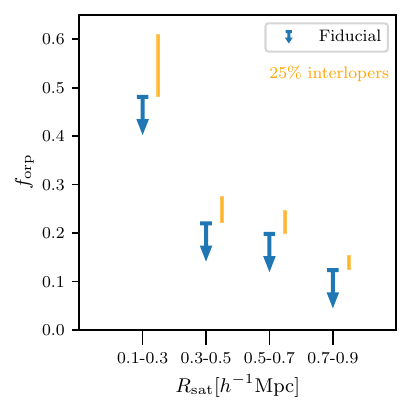}
    \caption{The figure shows the upper bound on the prevalence of orphan galaxies with cluster centric distance R$_{\rm Sat}$, by accounting the mass difference between the control sample galaxies and the satellites with orphan galaxies only.
    %The errorbars depict 16 and 84 percentile  confidence limit on this parameter.
    The yellow arrows indicate the shift in orphan fractions assuming some contamination from interloper field galaxies. HSC-observations suggest that, more satellites become orphans if they evolve at nearby distances from the BCG, i.e. in the strong gravitational influence of the central galaxy, in comparison to the satellites which evolve at far-away distances. }
 \label{fig:orphan_fraction}
\end{figure}

We note that the constraints derived above assume that the redMaPPer memberships to determine the satellite galaxies in galaxy clusters are reliable. Given that the cluster memberships in redMaPPer are determined only using photometric data, it is likely that the satellite galaxy classification suffers from contamination by field galaxies due to projection effects \citep[see e.g.,][]{Zu2017, Sunayama2020}. \citet{Myles_2021} use spectroscopic redshifts available for redMaPPer member galaxies in order to determine that a fraction $f_{\rm cont}=0.25$ of satellite galaxies identified by the algorithm may in reality be field galaxies projected along the line-of-sight. Such a contamination would result in an orphan fraction limit which gets relaxed by a factor $(1-f_{\rm cont})^{-1}$. In Fig.~\ref{fig:orphan_fraction}, we show the effect of such contamination by interlopers using the orange lines.

We further note that there is a possibility that the contamination is dependent upon the distance of the putative satellite galaxies from the cluster center, with smaller contamination in the inner regions. Therefore comparisons with simulations should be done with caution and should include possible contamination of the satellite galaxy sample due to projection. We defer such detailed comparisons to future work.

\section{Summary}
\label{sec:summary}
In this work we used the first year weak lensing catalog from the Subaru HSC survey to investigate the dark matter subhalos that surround satellite galaxies in galaxy clusters selected from the redMaPPer cluster catalog, within the redshift range $0.1\leq z \leq 0.33$. Our results can be summarized as follows:

\begin{itemize}
\item We measured the weak lensing signal of redMaPPer satellite galaxies binned in 4 projected cluster-centric distance bins $R\in (0.1, 0.3], (0.3, 0.5], (0.5, 0.7], (0.7, 0.9]$ with a total signal to noise ratio of 20.74, 18.00, 17.66, 17.35. 
\item We constructed a control sample of galaxies which do not reside in a cluster environment but have similar distribution of fluxes in the different optical wavelength bands as the satellite galaxy sample. We also measured the weak lensing signal around the control sample of galaxies to compare with that of the satellite galaxies.
\item We modelled the weak lensing signal both around satellites and control sample with the NFW profile, accounting for various systematic effects corresponding to miscentering of BCGs, the finite radial bin width as well as the radial dependence of the number of lens-source pairs. Our model provides a good description of the data and allows robust constraints on the masses of the subhalos of satellite galaxies and the masses of the control sample.
\item We find that the subhalo masses of satellite galaxies get systematically smaller than that of the control sample, with differences as large as factor of three in the innermost cluster-centric distance bin (although with large error given our current statistical precision). The subhalo masses approach that of the control sample as we go further from the cluster center, consistent with differences expected from tidal stripping of the dark matter subhalos.
\item We do not see a strong dependence of the average subhalo masses of redMaPPer satellite galaxies on the cluster-centric distances.
\item Attributing the difference between the subhalo masses of the satellites and those of the control sample to orphan galaxies which have completely lost their subhalos, allows us to obtain the first empirical upper limits on the fraction of orphan galaxies allowed at various cluster centric distances. We find an upper limit on the orphan fraction of about 48 percent in the innermost radial bin $r\in[0.1, 0.3]$ \mpch. These upper limits fall off to smaller values at larger galactocentric distances.
\end{itemize}

In our work we have only considered simple models for the subhalo density profiles given the precision of our current measurements \citep[see][~for the effects due to tidal truncation]{Sifon2015}. As the statistical precision increases, better models which include effects related to tidal truncation of the subhalos can also be used.

The Subaru HSC survey has recently published its 3 year catalog \citep{Li2022} based on roughly three times larger area than the first year. The data gathering operations from the entire HSC survey with about eight fold larger area have also concluded. Once the final year shape catalogs will be available, the lensing measurements around satellite galaxies will become more precise and will allow further detailed investigations. Albeit less significant with the current data, we do see signs that systematic effects will become further important as the statistical errors go down. We defer such investigations to future work.

\section*{Acknowledgements}
The authors would like to thank  Susmita Adhikari, Arka Banerjee, Souradip Bhattacharyya, Navin Chaurasiya, Priyanka Gawade, Preetish K Mishra, Aseem Paranjape, Masahiro Takada and  Crist{\'o}bal {Sif{\'o}n} for the useful discussions and suggestions throughout this work and on the draft version of this manuscript. We thank the referee Stella Seitz for her useful comments which helped us to further improve the analysis in this paper. The authors acknowledge the use of Pegasus, the high performance computing (HPC) facility of IUCAA.

The Hyper Suprime-Cam (HSC) collaboration includes the astronomical communities of Japan and Taiwan, and Princeton University. The HSC instrumentation and software were developed by
the National Astronomical Observatory of Japan (NAOJ), the Kavli
Institute for the Physics and Mathematics of the Universe (Kavli
IPMU), the University of Tokyo, the High Energy Accelerator Research Organization (KEK), the Academia Sinica Institute for Astronomy and Astrophysics in Taiwan (ASIAA), and Princeton University. Funding was contributed by the FIRST program from the
Japanese Cabinet Office, the Ministry of Education, Culture, Sports, Science and Technology (MEXT), the Japan Society for the Promotion of Science (JSPS), Japan Science and Technology Agency
(JST), the Toray Science Foundation, NAOJ, Kavli IPMU, KEK,
ASIAA, and Princeton University.

This paper is based [in part] on data collected at the Subaru Telescope and retrieved from the HSC data archive system, which is operated by Subaru Telescope and Astronomy Data Center (ADC) at NAOJ. Data analysis was in part carried out with the cooperation of Center for Computational Astrophysics (CfCA), NAOJ.

\section{Data Availability}
The satellite galaxy sample used in this analysis, constructed from publicly available redMaPPer\footnotemark \footnotetext{http://risa.stanford.edu/redmapper/} cluster/member catalog  with redshift and $P_{\rm cen}$ cuts, as described in  Section~\ref{sec:redmapper} along with the control sample of field galaxies corresponding to these satellites, can be accessed from repository \@ \url{https://github.com/ratewalamit/environmental\_effects.git}. This repository also contains  weak lensing measurements around the lens galaxies in  satellite and control sample catalogs, covariance matrices obtained using shape noise and the Jackknife technique, posterior distribution of the model parameters for all the bins used in the analysis.
\bibliographystyle{mnras}
\bibliography{bibliography} % if your bibtex file is called example.bib

%%%%%%%%%%%%%%%%% APPENDICES %%%%%%%%%%%%%%%%%%%%%
\appendix
\renewcommand{\thefigure}{A\arabic{figure}}
\setcounter{figure}{0}
\renewcommand{\thetable}{A\arabic{table}}
\setcounter{table}{0}

\section{Posterior distribution with Shape noise Covariance}
We also perform this whole analysis by using the shape noise covariance to estimate the posterior distributions using Eq.~\ref{eq:posterior}. In Table:\ref{table:parameter_comparsion}, we compare the results obtained from both of these covariance estimation techniques. Since both the techniques result in similar covariance values on scales used to infer the satellite subhalo masses and the halo masses of the control galaxy sample, we conclude that choice of the technique used to calculate covariance matrix does not strongly impact our conclusions.
\begin{table}
\centering 
 \renewcommand{\arraystretch}{1.5}
\rule{0pt}{4ex}
 \begin{tabularx}{\columnwidth} {    %fixing width of the table
   >{\centering\arraybackslash}X 
   >{\centering\arraybackslash}X 
   >{\centering\arraybackslash}X}
  \hline\hline
    Bin $[h^{-1}\rm Mpc$] & $f_{\rm orp}$(Jackknife)& $f_{\rm orp}$(Shape noise) \\
 \hline
    0.1-0.3 & 0.48 & 0.39\\ 
    0.3-0.5 & 0.22 &  0.16  \\
    0.5-0.7 &  0.20 & 0.19 \\
    0.7-0.9 &  0.12 & 0.10\\
    [0.5ex]
\hline
\end{tabularx}
\caption{In this table we compare the constraints on the orphan fraction at different distances from the cluster center, as obtained by using different covariance estimation techniques, i.e. jackknife versus shape noise. We see the results from both techniques are consistent within errors.}
\label{table:parameter_comparsion}
\end{table}

\begin{table*}
 \centering
 \renewcommand{\arraystretch}{1.5}
 \begin{tabularx}{\textwidth} {    %fixing width of the table
  | >{\centering\arraybackslash}X 
| >{\centering\arraybackslash}X 
  | >{\centering\arraybackslash}X  
  | >{\centering\arraybackslash}X 
  | >{\centering\arraybackslash}X 
  | >{\centering\arraybackslash}X 
  | >{\centering\arraybackslash}X | }
\hline
  $\rm R_{sat} ~$[$h^{-1}\rm Mpc$]& $ \rm \log \dfrac{M_{\rm clu}}{[~h^{-1}M_{\odot}]}$ & $ \rm \log \dfrac{M_{\rm sat}}{[~h^{-1}M_{\odot}]}$ & $ \rm \log \dfrac{M_{\rm cont}}{[~h^{-1}M_{\odot}]}$ & $ \rm \log \dfrac{M_{\rm bary}}{[~h^{-1}M_{\odot}]}$ &  $\sigma_{\rm mc} [h^{-1}\rm Mpc]$\Tstrut\Bstrut\\
 \hline\hline
  (0.1-0.3] & $14.35^{+0.02}_{-0.02}$ & $11.91^{+0.19}_{-0.20}$ & $12.24^{+0.06}_{-0.07}$ & $10.56^{+0.15}_{-0.25}$ & $0.49^{+0.36}_{-0.35}$\\
 \hline
(0.3-0.5] & $14.36^{+0.02}_{-0.02}$ & $12.19^{+0.08}_{-0.10}$ & $12.29^{+0.06}_{-0.06}$ & $9.93^{+0.40}_{-0.58}$ & $0.49^{+0.33}_{-0.29}$\\
 \hline
(0.5-0.7]  & $14.40^{+0.02}_{-0.02}$ & $12.01^{+0.10}_{-0.14}$ & $12.13^{+0.07}_{-0.08}$ & $10.57^{+0.15}_{-0.24}$ & $0.49^{+0.33}_{-0.30}$\\
 \hline
 (0.7-0.9]  & $14.43^{+0.02}_{-0.02}$ & $12.15^{+0.07}_{-0.09}$ & $12.21^{+0.06}_{-0.07}$ & $10.60^{+0.13}_{-0.21}$ & $0.48^{+0.33}_{-0.31}$\\[0.5ex]\hline
\end{tabularx}
\begin{tabularx}{\textwidth} {    %fixing width of the table
  | >{\centering\arraybackslash}X 
  | >{\centering\arraybackslash}X 
  | >{\centering\arraybackslash}X 
  | >{\centering\arraybackslash}X  
  | >{\centering\arraybackslash}X 
  | >{\centering\arraybackslash}X 
  | >{\centering\arraybackslash}X | }
\hline
  $\rm R_{sat} ~$[$h^{-1}\rm Mpc$]& $ \rm \log \dfrac{M_{\rm clu}}{[~h^{-1}M_{\odot}]}$ & $ \rm \log \dfrac{M_{\rm sat}}{[~h^{-1}M_{\odot}]}$ & $ \rm \log \dfrac{M_{\rm bary}}{[~h^{-1}M_{\odot}]}$ &  $\sigma_{\rm mc} [h^{-1}\rm Mpc]$ &$f_{\rm orp}$\Tstrut\Bstrut\\
 \hline\hline
  (0.1-0.3] & $14.34^{+0.02}_{-0.02}$ & $12.23^{+0.06}_{-0.07}$ & $10.57^{+0.15}_{-0.24}$ & $0.50^{+0.35}_{-0.35}$ & 0.39\\
 \hline
(0.3-0.5] & $14.36^{+0.02}_{-0.02}$ & $12.29^{+0.06}_{-0.06}$ & $9.95^{+0.39}_{-0.58}$ & $0.49^{+0.34}_{-0.29}$& 0.16\\
 \hline
(0.5-0.7]  & $14.40^{+0.02}_{-0.02}$ & $12.13^{+0.07}_{-0.08}$ & $10.58^{+0.14}_{-0.23}$ & $0.48^{+0.34}_{-0.30}$ & 0.19\\
 \hline
 (0.7-0.9]  & $14.43^{+0.02}_{-0.02}$ & $12.21^{+0.06}_{-0.07}$ & $10.61^{+0.13}_{-0.20}$ & $0.48^{+0.33}_{-0.31}$ & 0.10\\[0.5ex]\hline
\end{tabularx}
\caption{Similar to Table. \ref{table:inferred_parameters}, but using shape noise covariance matrices.}
\label{table:inferred_parameters_shapenoise}
\end{table*}

%%%%%%%%%%%%%%%%%%%%%%%%%%%%%%%%%%%%%%%%%%%%%%%%%
\bsp	% typesetting comment
\label{lastpage}
\end{document}